\documentclass{andp2012}
\usepackage[english]{babel}
\usepackage{relsize}
\usepackage{graphicx,float}
\usepackage{epsfig,booktabs}
\usepackage{times}
\usepackage{wrapfig}
\usepackage[utf8]{inputenc}
\usepackage{amsmath,amssymb,amsfonts,amsthm,epstopdf}
\usepackage{nccmath}
\usepackage{lineno,hyperref,amssymb,caption,subcaption,graphicx,float}
 
\pdfoutput=1

\category{Review Article}

\title{Electronic Transport and Thermopower in 2D and 3D Heterostructures--A Theory Perspective}

\author[A.\,K. Majee]{Arnab K. Majee\inst{1}}
\author[A. Kommini]{Adithya Kommini\inst{1}}
\author[Z. Aksamija]{Zlatan Aksamija\inst{1}\footnote{Corresponding author\quad E-mail:~\textsf{zlatana@engin.umass.edu}}}
\address[1]{Electrical and Computer Engineering Department, University of Massachusetts Amherst, Amherst USA}
\shortauthors{Majee et al.}

\begin{abstract}
	In this review, we discuss the impact of interfaces and heterojuctions on the electronic and thermoelectric transport properties of materials. We review recent progress in understanding electronic transport in heterostructures of two-dimensional (2D) materials ranging from graphene to transition metal dichalcogenides (TMDs), their homojunctions (grain boundaries), lateral heterojunctions (such as graphene/MoS$_2$ lateral interfaces), and vertical van der Waals (vdW) heterostructures. We also review work on thermopower in 2D heterojunctions, as well as their applications in creating devices such as resonant tunneling diodes (RTDs). Lastly, we turn our focus to work in three-dimensional (3D) heterostructures. While transport in 3D heterostructures has been researched for several decades, here we review recent progress in theory and simulation of quantum effects on transport via the Wigner and non-equilibrium Green's functions (NEGF) approaches. These simulation techniques have been successfully applied toward understanding the impact of heterojunctions on transport properties and thermopower, which finds applications in energy harvesting, and electron resonant tunneling, with applications in RTDs. We conclude that tremendous progress has been made in both simulation and experiments toward the goal of understanding transport in heterostructures and this progress will soon be parlayed into improved energy converters and quantum nanoelectronic devices.

\end{abstract}

\begin{document}
	
\maketitle

\section{Introduction and overview}

Progress in understanding charge transport in two-dimensional (2D) materials is forging along at a furious pace. It follows an arc paralleling that of three-dimensional (3D) materials a few decades prior: after their intrinsic (bulk) properties were well studied, researchers turned to enhancing the intrinsic characteristics by combining materials together into heterostructures. This approach brings both tremendous complexity and opens up vistas to emerging properties not available in individual materials or atomic monolayers. Such heterostructures have device applications ranging from CMOS transistors with few-nanometer channel lengths that continue to track the scaling predictions of Moore's law to beyond-CMOS devices such as thermoelectric energy converters and resonant tunneling diodes based on quantum tunneling. The common thread across all the aforementioned materials and applications is that their functionality revolves around electron transport across interfaces between dissimilar materials, called heterojunctions. The resulting heterostructures alter the transmission of electrons and lead to increased contributions from tunneling, resonant quantum states, or energetic electrons with improved thermoelectric coefficients. 

In order to cover the recent progress, while at the same time drawing parallels between 2D and 3D heterostructures, we organize the paper as follows: first we cover 2D materials, starting with a brief history before delving into graphene and MoS$_2$ grain boundaries. Next, we review lateral heterostructures, such as those formed between graphene and MoS$_2$ in Sec.~\ref{sec:lateral}, and include band alignment theory and resonant tunneling diode (RTD) applications. We round out the coverage of 2D materials with vertical (stacked) heterostructures in Sec.~\ref{sec:vertical}, including progress on applications in thermoelectric energy conversion. Then we switch over to 3D heterostructures and periodic superlattices in Sec.~\ref{sec:3Dhetero}, covering the history of tunneling diodes, the theory and numerical simulation methods for quantum transport in heterostructures, and their applications in enhancing thermoelectric energy conversion efficiency of semiconductor materials. We end with conclusions and future outlook.

\section{Charge transport in two-dimensional materials}

Although the idea of graphene (a single layer of graphite) existed theoretically for several decades, it was believed that crystalline two-dimensional (2D) materials, including graphene, could not exist in nature because of thermodynamical instability~\cite{PeierlsAnnales1935}. It was in 2004, when a group of scientists from University of Manchester in United Kingdom and Institute for Microelectronics Technology in Russia successfully separated a single layer from bulk graphite for the first time~\cite{NovoselovScience04}. Their groundbreaking work resulted in the 2010 Nobel Prize in Physics being awarded to Novoselov and Geim. They followed it up by preparing single-layer exfoliated forms of other van der Waals 2D atomic crystals, including transition metal dichalcogenides (TMDs) such as MoS$_2$ and NbSe$_2$, and boron nitride~\cite{NovoselovPNAS05}. This cohort of 2D materials includes graphene, hexagonal boron nitride (hBN), and transition metal dichalcogenides (TMDs), which have covalent in-plane bonds and weak van der Waals (vdW) bonds across atomic planes. The weak interlayer coupling allows these vdW monolayers to be separated, manipulated, and assembled into vertical and lateral heterostructures. These materials have a host of unique and superlative electronic, optical, thermal, and thermoelectric properties, thus forming a basis for future low-power logic devices, high-efficiency energy materials, and high-performance optoelectronics.

The unique electronic structure of graphene, where the low energy charged carriers mimic relativistic particles with zero rest mass having an 'effective speed of light' of 10$^6$ m s$^{-1}$~\cite{NovoselovNature05}, along with experimental demonstrations of integer quantum Hall states~\cite{ZhangNature05} and Klein tunneling~\cite{GeimNatureMater07}, not only opened up avenues for rich, fundamental science but also led to a possibility of realizing new electronic and magneto-electronic device applications. Earlier reports suggested that carrier mobilities, which were found to be about 15,000 cm$^{2}$ V$^{-1}$ s$^{-1}$ for both electron and holes, were independent of temperature over the range between 10 and 1000 K~\cite{NovoselovNature05}, suggesting a very strong influence from the interaction with the underlying substrate. This has led to a series of studies trying to establish the upper bound of the intrinsic phonon-limited mobility by fabricating and measuring highly crystalline, suspended graphene~\cite{BolotinSolidStateCommun08,BolotinPRL08,DuNatNano08}. 

An electron mobility of 230,000 $\text{cm}^{\text{2}}~\text{V}^{\text{-1}}~\text{s}^{\text{-1}}$ at a carrier concentration of $2\times10^{11}$ cm$^{-2}$ is among the highest ever measured mobility in suspended graphene~\cite{BolotinSolidStateCommun08}. This value can be considered as the phonon-limited upper bound \cite{SuleJAP12}. Despite such high mobilities, the weak electrostatic gating and absence of band gap makes it an undesirable candidate for 2D field-effect transistor (FET). However, it did not deter scientific community from continuing research on graphene because of its rich and unique fundamental physical properties. On the other hand, the inherent band gap of TMDs made them promising candidates for making FETs. So, research turned to this new family of 2D materials based on TMDs. Early studies of MoS$_2$ found significantly lower mobility but with a much better electrostatic control. The subsequent introduction of a substrate~\cite{ChenNatNano08,NagashioJAP11,DeanNatNano10,LiNanoLett09} for better electrostatic gating provided an opportunity to measure large-area graphene sheets, although substrate impurity \cite{SulePRB14}, interface, and remote phonon \cite{OngPRB12} scattering significantly reduce carrier mobility.

\begin{figure}[ht!]
	\centering
	\includegraphics[width=0.8\columnwidth]{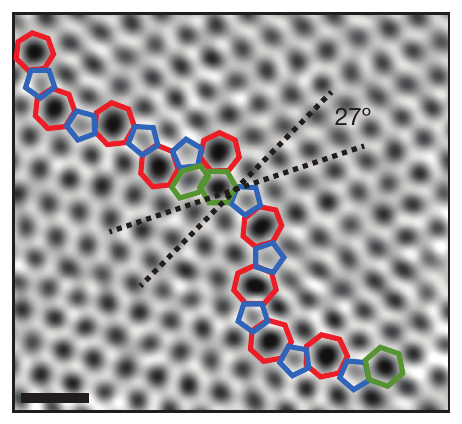}
	\caption{shows atomic resolution ADF-STEM image of two graphene grains stitched together to form GB with a misorientation angle of 27$^\circ$. GB is composed of line defects made up of periodic pentagonal and heptagonal rings. Reprinted by permission from Nature Publishing Group~\cite{HuangNature11}, Copyright(2015).}
	\label{fig:figure3c}
	\label{overflow}
\end{figure} 

In the subsequent years, researchers developed improved techniques for growing large-area samples of single crystalline graphene. Hao et al.~\cite{HaoScience13} demonstrated reproducible growth of centimeter-sized single-crystalline graphene grains. This paved the way for wafer-scale growth of single crystalline domains of graphene~\cite{LeeSCience14} and MoS$_2$~\cite{YuACSnano17} in a lab environment. However, for commercial purposes, cost-effectiveness and high-throughput synthesis of wafer-scale single-crystalline 2D materials are the major requirements to compete with the existing silicon technology. To date, chemical vapor deposition (CVD) is the cheapest and fastest method of growing such large-area sheets of 2D materials, ~\cite{LiScience09,ReinaNanoLett09,KimNature09,JuangCarbon10} but CVD-grown sheets are inherently polycrystalline in nature. The grains grow from each metal seed point until the point where they meet neighboring grains, where they form grain boundaries (GBs). Each grain is randomly oriented in space, and thus has different crystallographic orientation with angles varying between 0$^\circ$ to 60$^\circ$. Different crystallographic orientation of adjacent grains causes a mismatch in the crystal structure at their boundary, leading to the formation of pentagon-heptagon pairs \cite{YazyevPRB10,HuangNature11}. Macroscopically, the relative mismatch between the crystallographic orientations of adjacent grains is referred to misorientation angle. Figure~\ref{fig:figure3c} shows atomic-resolution ADF-STEM images of graphene crystals and a GB with misorientation angle of 27$^\circ$\cite{HuangNature11}. To realize practical applications of large-area 2D devices made up of graphene or TMDs, we believe that it is crucial to develop a detailed understanding of the impact of GB misorientation angles on electrical transport in polycrystalline sheets. To this end, numerous experimental ~\cite{LiNanoLett10,MaNatCommun17,HuangNature11} and theoretical studies ~\cite{VancsoCarbon13,SunRSCadv16} have been carried out.
 
 \begin{figure*}[t]
 	\centering
 	\begin{minipage}[t][6cm][c]{0.51\textwidth}
 		\begin{center}
 			\includegraphics[scale=0.42]{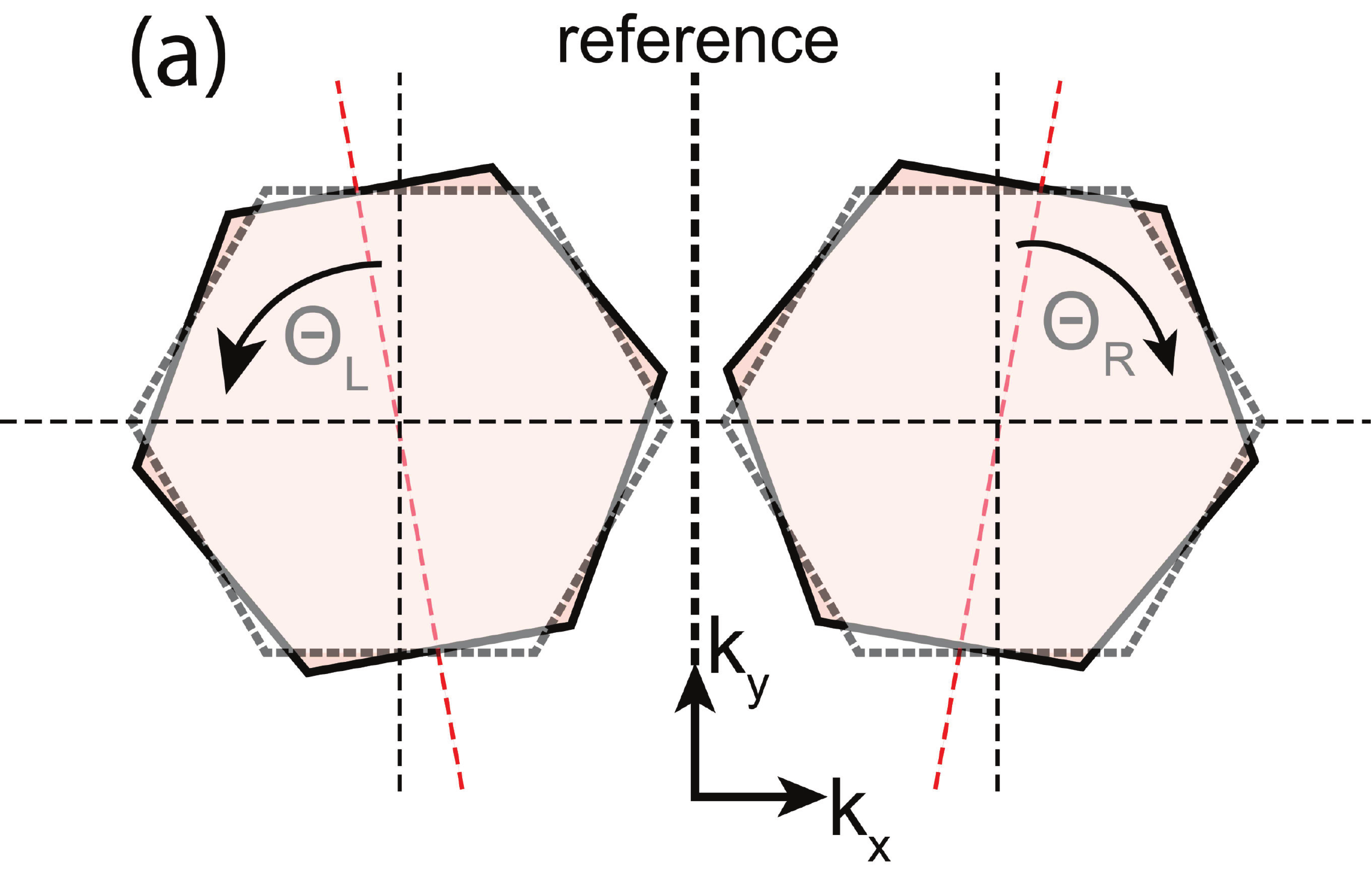}
 		\end{center}
 		\label{fig:aaa}
 	\end{minipage}
 	\centering
 	\begin{minipage}[t][6cm][c]{0.4\textwidth}
 		\begin{center}
 			\includegraphics[scale=0.39]{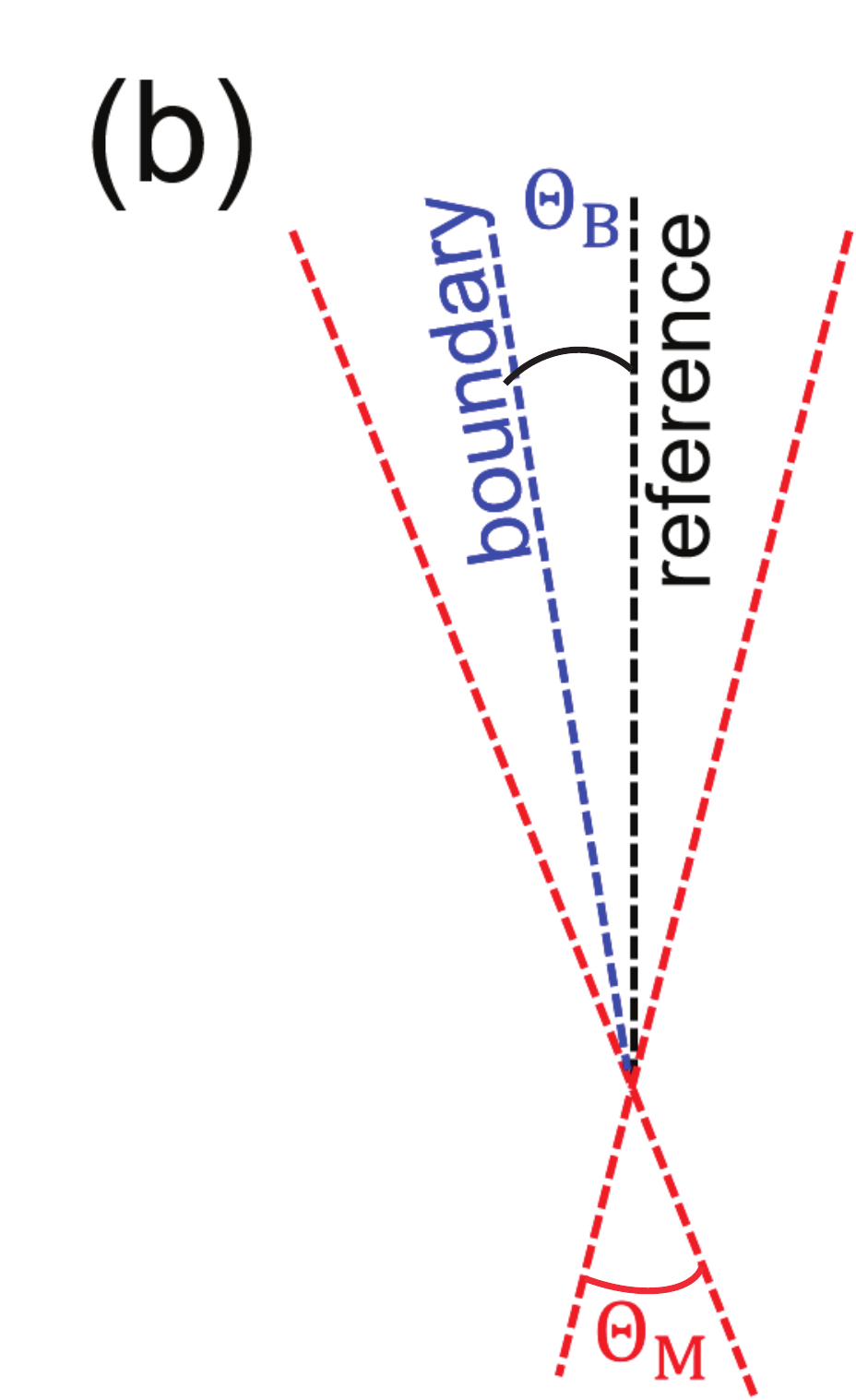}
 		\end{center}
 		\label{fig:bandstructure}
 	\end{minipage}
 	\caption{\textbf{(a)} shows crystallographic orientations of the grains with respect to the interface. The dashed-hexagons represent the BZs for perfectly matched condition ($\Theta_L=\Theta_R=0^\circ$). $\Theta_L$ is the angle of rotation, measured in anticlockwise direction, between the rotated left grain (solid-outlined hexagon) and the one for perfectly-matched condition (dash-outlined hexagon). $\Theta_R$ is the angle of rotation measured in clockwise direction, and the total misorientation angle is $\Theta_M=\Theta_L + \Theta_R$. \textbf{(b)} shows the boundary angle $\Theta_B$ with respect to the reference for a given $\Theta_M$. These figures are adapted from [\citenum{MajeeSciRep17}].}
 	\label{fig:figure1}
 	\label{overflow}
 \end{figure*}
 
\subsection{Charge transport across graphene grain boundaries} 
The earliest studies focused electrical resistance across graphene grain boundaries. Like GBs in bulk semiconductors, GBs in graphene were also found to impede electrical conduction. Experimental studies show that the resistivity of graphene GBs varies over a broad range from few $\Omega$ $\mu \text{m}$ ~\cite{GrosseApplPhysLett14} to several tens of k$\Omega$ $\mu \text{m}$ ~\cite{YuNatMater11,JaureguiSolidStateCommun11}. There are many factors that are responsible for this large variation of resistivity. These include grain misorientation angles ~\cite{ZhangJPhysChemC14}, GB roughness, wrinkles and transition width \cite{AhmadNanotech12}, impurities and structural defects at the boundaries~\cite{TapasztaAPL12, FeiNatNano13, KoepkeACSnano2013, NemesInczeCarbon13}, position of the Fermi level, and other differences arising due to processing and measurement techniques. Most of the time, it is challenging to isolate the effect of one of these factors from others. To minimize the large spread in resistivity, there has been a constant effort on improving device quality ~\cite{SunNature10,WuApplMaterInterfaces12,SunCarbon16,XuNanotechnology17,KangNanoscale12,BaeNatureNano10} by minimizing impurities and structural defects at the boundaries. Even in high quality large-area samples, the effect of misorientation angles on charge transport is quite significant. 

Figure~\ref{fig:figure1}(a) shows a schematic of the Brillouin zones (BZs) of two adjacent grains. BZ of the left grain is rotated by $\Theta_L$ in the anti-clockwise direction, whereas the one on the right by $\Theta_R$ in the clockwise direction such that the relative misorientation is given by $\Theta_M=\Theta_L+\Theta_R$. $\Theta_B$ is the angle that boundary (blue-dashed line) makes with respect to the reference, shown in the Figure~\ref{fig:figure1}(b), (the hexagons are omitted for clarity and the red-dashed lines represent the crystallographic orientation of each grain). For a given misorientation angle $\Theta_M$, the type of GB is decided based on $\Theta_B$. For example, when $\Theta_B=0^\circ$ GB is symmetric, and is called \emph{twin} GB. On increasing $\Theta_B$, that is by rotating the boundary towards the crystallographic direction of one of the grains, the GB becomes asymmetric, and is called \emph{tilt} GB. Finally, when the boundary completely aligns with the crystallographic direction of the left/right grain, that is $\Theta_B=\Theta_M/2$, it is the most asymmetric GB that can be obtained for a given $\Theta_M$.

\begin{figure}[ht!]
	\centering
	\includegraphics[width=0.9\columnwidth]{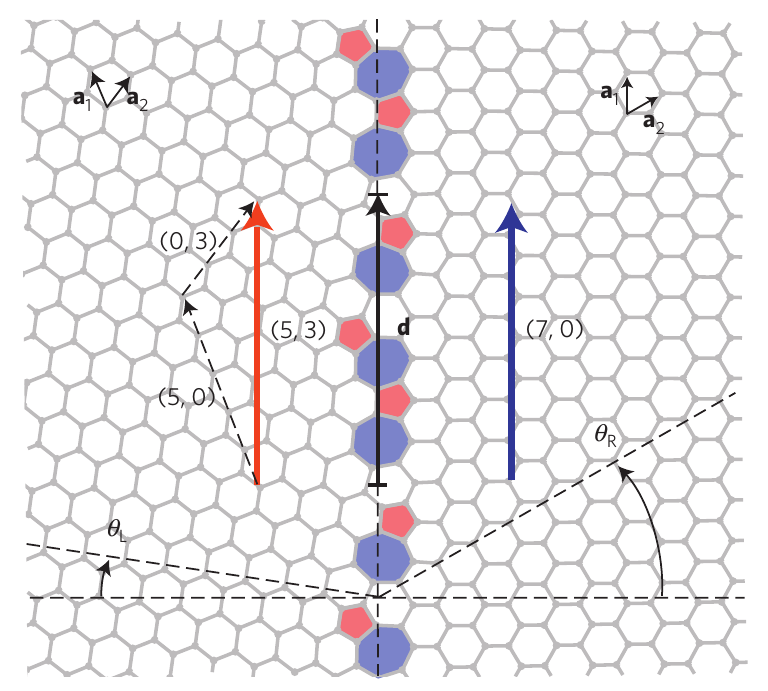}
	\caption{shows an example of graphene tilt GB with misorientation angle of 38.2$^\circ$ ($\Theta_L$=8.2$^\circ$ and $\Theta$=30$^\circ$) obtained from the dislocation model described by\cite{YazyevPRB10}. Reprinted by permission from Nature Publishing Group~\cite{YazyevNatMater10}, Copyright(2010).}
	\label{fig:figure3b}
	\label{overflow}
\end{figure}

The transport of carriers across any GB requires quantum-mechanical wave continuity~\cite{ChenBook05}, based on which the energy as well as transverse momentum (momentum parallel to the GB/interface) of the incident carrier should be conserved on transmission. Mathematically, for every incident carrier $k_i$ we find a transmitted wavevector $k_t$ such that $k_{i_{\parallel}}=k_{t_{\parallel}}$ and $E(k_i)=E(k_{t_{\parallel}}+k_{t_{\perp}})$, where $k_{i_{\parallel}}$ and $k_{t_{\parallel}}$ are the components of incident and transmitted wavevectors respectively along the GB direction.  $k_{i_{\perp}}$ and $k_{t_{\perp}}$ are the components of incident and transmitted wavevectors respectively perpendicular to the GB direction. This is analogous to the Acoustic Mismatch model in phonon transport where the transmission across a GB is considered to be specular, that is the effect of GB roughness is negligible. Then the mode-dependent transmission coefficient $\tau_b (k_i)$ for every $k_i$ per branch $b$ is calculated by
\begin{ceqn}
	\begin{align}\label{eq:ModeTransmission}
	\tau_b (k_i)=\frac{\mid4k_{i\perp} k_{t\perp}\mid}{\mid k_{i\perp}+k_{t\perp}\mid^2}
	\end{align}
\end{ceqn}
Next, $\tau_b (k_i)$ is converted to energy-resolved transmission coefficient $\Gamma_b (E)$ by averaging over the contour with constant energy, described by $\delta(E-E_b (k_i))$. This approach uses the 2D version of the linear extrapolation described by Gilat and Raubenheimer\cite{GilatPhysRev66} as
\begin{ceqn}
	\begin{align}\label{eq:EnergyTransmission}
	\Gamma_b (E)=\frac{1/4\pi^2 \int \tau_b (k_i) \delta(E-E_b (k_i)) dk_i}{1/4\pi^2 \int \delta(E-E_b (k_i)) dk_i}
	\end{align}
\end{ceqn}
The denominator in Eq. (\ref{eq:EnergyTransmission}) is the density of states per branch $D_b(E)$. The transport distribution function TDF can then be calculated as
\begin{ceqn}
	\begin{equation}\label{eq:TDF}
	\Xi(E)=\sum_{b}^{}v_b(E) \Gamma_b(E) D_b(E)
	\end{equation}
\end{ceqn}
where mode-dependent velocity in the direction of transport $v_{b_\parallel}(k_i)$ is converted to energy-resolved velocity $v_{b_\parallel}(E)$ in the same way as transmission coefficient. The TDF is then used to numerically compute GB conductance in the Landauer Formalism as follows
\begin{ceqn}
	\begin{equation}\label{eq:RGB}
	G_{GB/int.}=\frac{e^2}{2} \mathop{\mathlarger{\int_{E_{C}}^{E_{max}}}} \Xi(E)\left(-\frac{\partial f(E-E_F,T)}{\partial E}\right)dE
	\end{equation}
\end{ceqn}
$f(E)=[1+exp((E-E_F)/k_BT)]^{-1}$ is the equilibrium Fermi-Dirac distribution function. In the past, models have also been developed to study the microscopic effects---atomic structure\cite{YazyevNatMater10} and buckling \cite{DechampsNanoscale18}---on GB resistance, where periodic GBs are constructed using a dislocation model described by \cite{YazyevPRB10}. An asymmetric GB with $\Theta_L$ and $\Theta_R$ equal to 8.2$^\circ$ and 30$^\circ$ respectively, constructed from the dislocation model is shown in Figure~\ref{fig:figure3b}. The quantum transmission and electronic transport properties were calculated from first principle Density Functional Theory (DFT) using a non-equilibrium Green's function (NEGF) formalism implemented in the TRANSIESTA code\cite{BrandbygePRB02}.

Clark et al. ~\cite{ClarkACSnano13} measured GB resistivity $R_{GB}$ and found that there exists a positive correlation between $R_{GB}$ and misorientation angles $\Theta_M$, but the effect on resistivity due to a GB transition width could not be separated. In a recent study, we demonstrated that, although $\Theta_M$ plays an important role in determining $R_{GB}$, it is the relative angle of adjacent grains with respect to the GB that has more influence on $R_{GB}$ than $\Theta_M$ alone ~\cite{MajeeSciRep17}. A GB formed between a pair of grains which can be represented by same translational vectors (component of an unit vector representing crystal orientation of each grain along GB direction) are found to be highly conductive (about 80$ \%$), whereas the ones with different translational vectors behave as reflectors of carriers ~\cite{YazyevNatMater10}. In other words, symmetric GBs are highly transparent to carriers in intrinsic (defect-free) graphene ~\cite{ZhangJPhysChemC14}.

\begin{figure*}[ht!]
	\centering
	\includegraphics[width=6.5in]{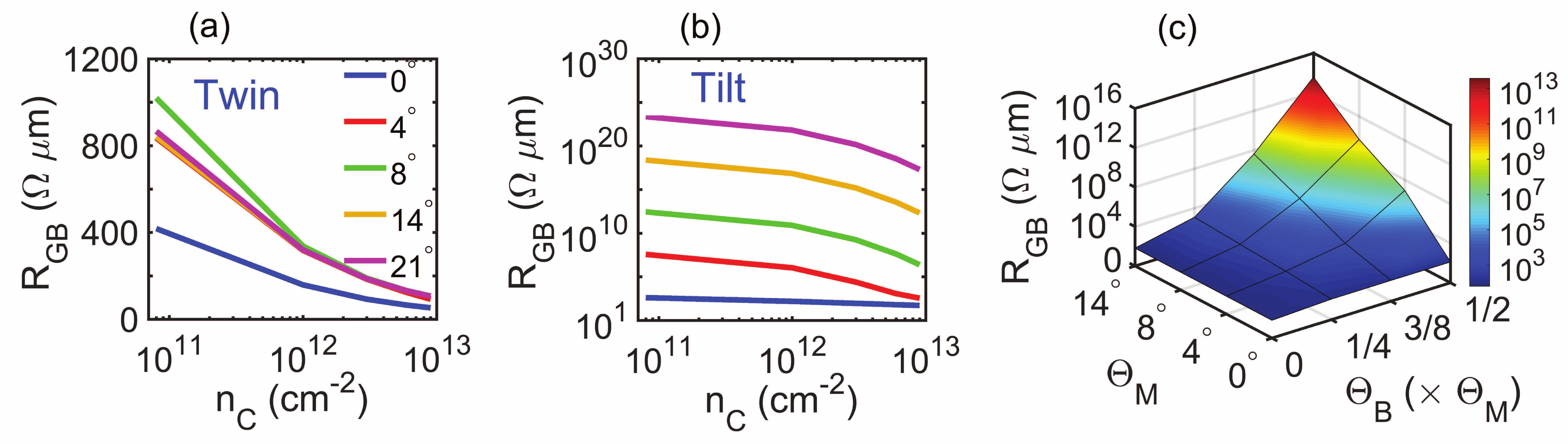}
	\caption{Resistivities across graphene GBs have been plotted against carrier concentration for different misorientation angles in \textbf{(a)} twin GBs ($\Theta_L=\Theta_R=\Theta_M/2$) \textbf{(b)} the most asymmetric GB ($\Theta_L=0^\circ $ and $ \Theta_R=\Theta_M$). The curves for large mismatch angles ($14^\circ$ and $21^\circ$) are overlapping on each other in \textbf{(a)}. \textbf{(c)} shows a surface plot of GB resistivity against $\Theta_M$ and $\Theta_B$. Here $\Theta_B$ is expressed in terms of $\Theta_M$. These figures are adapted from [\citenum{MajeeSciRep17}].}
	\label{fig:figure2}
	\label{overflow}
\end{figure*}  

Figure \ref{fig:figure2}(a) and \ref{fig:figure2}(b) show the dependence of GB resistance with carrier density for various misorientation angles in twin and tilt GBs, respectively. Several approaches have been employed to dope 2D materials which include diffusion or implantation of metals\cite{FangNanoLett13} and physi-/chemisorption of gases\cite{FangNanoLett12}, but direct doping has been found to be challenging in 2D materials. Consequently, modulation of carrier density in 2D FETs is mostly achieved by electrostatic gating. The typical carrier densities obtained from gating vary between 10$^{12}$ to 10$^{13}$ cm$^{-2}$. We found that GB resistance in twin GBs show a sharp dependence on carrier density, reaching the ballistic resistance value for n$_C$ equals 10$^{13}$ cm$^{-2}$. For tilt GBs, the resistance shows a decreasing trend even beyond 10$^{13}$ cm$^{-2}$ due to the conical bandstructure of graphene.

The effect of misorientation angles on GB resistance is illustrated in Figure \ref{fig:figure2}. Figure \ref{fig:figure2}(a) shows a weak dependence of $R_{GB}$ on misorientation angle in symmetric GBs, whereas in a completely asymmetric GB, the resistivity varies over 20 orders of magnitude with misorientation angles shown in Figure \ref{fig:figure2}(b). Symmetric rotation of BZ in twin GBs still conserve parallel momentum, and hence misorientation angles have negligible effect on GB resistivity. For tilt GBs, a fraction of the carriers cannot conserve parallel momentum, and hence get reflected which results in increase in GB resistivity. As graphene has a very steep and conical bandstructure near Dirac point (group velocity of about $10^5$ ms$^{-1}$), even a small rotation in BZs results in a large fraction of wavevectors which cannot conserve energy across GBs. This appears as an opening in the energy spectrum where there is no transmission, often referred to as \emph{transport gap}. Figure \ref{fig:figure2}(c) clearly shows that for a given misorientation angle there is a wide spread of GB resistivity depending on boundary angle $\Theta_B$. 
Despite numerous studies on graphene GBs, we believe that a systematic experimental demonstration of the impact of misorientation angle on electrical transport in clean (defect-free) samples is still missing.      

\subsection{MoS$_2$ Grain Boundaries}
\begin{figure}[ht!]
	\centering
	\includegraphics[width=0.8\columnwidth]{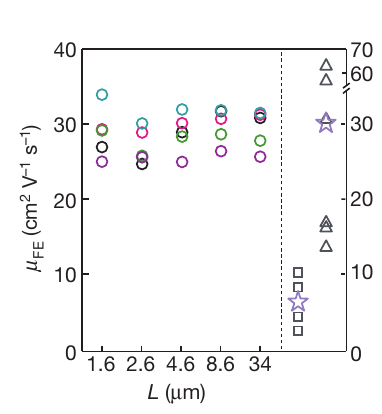}
	\caption{shows the field-effect mobility of MoS$_2$ FETs for five different lengths (ellipse markers). Data from literature on CVD-grown samples (squares) and exfoliated samples (triangles) are given on the right of the dotted line. Their medians are shown with purple stars. Reprinted by permission from Nature Publishing Group~\cite{KangNature15}, Copyright(2015).}
	\label{fig:figure3a}
	\label{overflow}
\end{figure}

Owing to their intrinsic bandgap, TMDs and especially MoS$_2$, have received significant research attention in recent years for electronic applications ~\cite{NajmaeiNatMater13,LyNatCommun16}. Although intrinsic room-temperature (RT) carrier mobilities in bulk and monolayer MoS$_2$ have been theoretically predicted to be about 400 and 1000 cm$^2$ V$^{-1}$ s$^{-1}$ respectively~\cite{FivazPR67,MaPRX14}, earlier studies reported a very low RT-carrier mobility of about 3 cm$^2$ V$^{-1}$ s$^{-1}$~\cite{NovoselovPNAS05}. This was attributed to the dominant charged impurity scattering and trapped charges at the substrate interface. Over the years, there have been significant efforts on fabricating ultra-clean samples, which led to a gradual improvement in RT-carrier mobilities~\cite{RadisavljevicNatNano11} upto more than 500 cm$^2$ V$^{-1}$ s$^{-1}$ in bulk samples of MoS$_2$ ~\cite{LiuIEEE12}. Beside the effect of surface impurities and trapped charges in single crystalline MoS$_2$, there had been a significant research dedicated to understand the effect of GBs in CVD-grown polycrystalline sheets. The field-effect mobility has been found to depend weakly on the channel length \cite{NajmaeiACSnano14,KangNature15} for lengths up to several tens of microns in CVD-grown MoS$_2$, shown in Figure~\ref{fig:figure3a}. Moreover, electrical properties of CVD-grown monolayer MoS$_2$ has been reported to be comparable to that of their exfoliated counterparts \cite{SchmidtNanolett14}. These studies indicate that GBs do not play a significant role in determining electrical properties in polycrystalline MoS$_2$ samples \cite{ZandeNatMater13}. Despite several studies on electrical properties of CVD-grown MoS$_2$, there is little known about the impact of misorientation angle on charge transport in CVD-grown MoS$_2$ sheets. 

\begin{figure*}[ht!]
	\centering
	\includegraphics[width=6.5in]{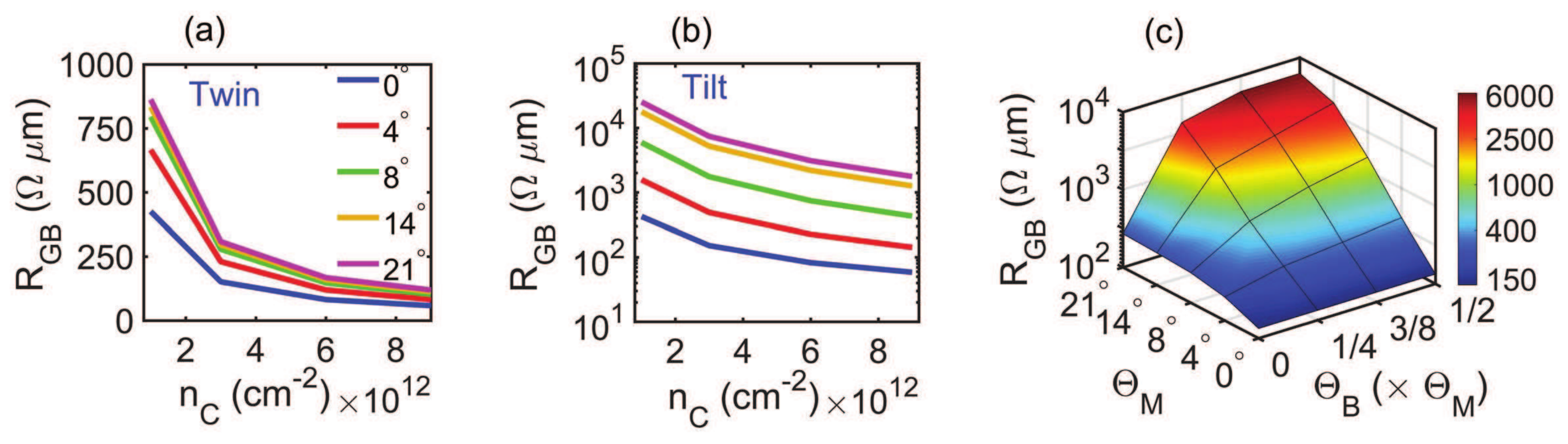}
	\caption{Resistivities across MoS$_2$ GBs have been plotted against carrier concentration for different misorientation angles in \textbf{(a)} twin GBs ($\Theta_L=\Theta_R=\Theta_M/2$) \textbf{(b)} the most asymmetric GB ($\Theta_L=0^\circ$ and $\Theta_R=\Theta_M$). The curves for large mismatch angles ($14^\circ$ and $21^\circ$) are overlapping on each other in \textbf{(a)}. \textbf{(c)} shows a surface plot of GB resistivity against $\Theta_M$ and $\Theta_B$. Here $\Theta_B$ is expressed in terms of $\Theta_M$. These figures are taken from [\citenum{MajeeSciRep17}].}
	\label{fig:figure3}
	\label{overflow}
\end{figure*}

In our recent study \cite{MajeeSciRep17}, we explained why GB resistivity is weakly dependent on misorientation angle in both twin and tilt MoS$_2$ GBs. In Figure \ref{fig:figure3}(a), it can be seen that GB resistivity of twin GBs vs. misorientation angle is comparable to that of twin graphene GBs. Figure \ref{fig:figure3}(b) shows that the resistivity of the most asymmetric-tilt GBs in MoS$_2$ varies over two orders of magnitude as compared to 20 orders of magnitude variation of resistivity in graphene tilt GBs with misorientation angle. We attribute this to the flatter parabolic conduction band (CB) of MoS$_2$. Even for large misorientation angles caused by the rotation of BZs of grains on either side of GB, there is a significant overlap in their CBs. As a result, a larger fraction of wavevectors conserve energy as compared to that in graphene. Consequently, the boundary angle parameterized by $\Theta_B$ does not largely influence GB resistivity in MoS$_2$ GBs  for a given misorientation angle $\Theta_M$, as can be seen in Figure \ref{fig:figure3}(c). It has been attributed to the flat parabolic bandstructure of MoS$_2$, in stark contrast to the steep cone-shaped bandstructure near the $K$-point in graphene. The density of states for MoS$_2$ is step-like, which makes it difficult to move the Fermi level deeper inside the CB by gating. Thus, GB resistance has been plotted for the typical range of carrier densities---10$^{12}$ to 10$^{13}$ cm$^{-2}$.

\section{Lateral heterojunctions in two-dimensional materials}\label{sec:lateral}

Owing to the potential of realizing all-2D devices including transistors, p-n junctions ~\cite{ChenChemMater16,LiACSnano16}, superlattices ~\cite{ZhangScience17}, tunneling devices ~\cite{LiIEDM16,MarinNanoscale17}, and lateral heterojunctions have drawn significant research interest in recent years. Few popular 2D lateral heterojunctions include graphene/TMD ~\cite{MajeeSciRep17,BehranginiaSmall17,Chen2DMater17,SunCompMaterSci17} TMD/TMD~\cite{ChenChemMater16,HuangNatureMat14,DuanNatureNano14}, and graphene/non-TMD heterojunctions ~\cite{SunRSCadv15,ChungJPhysD17,KiralyChemMater15,CiNatMater10,LiuNatNano13}. A recent review~\cite{LiMaterToday16} has tabulated different applications of various types of 2D heterojunctions. Here we are restricting ourselves to graphene/TMD and TMD/TMD junctions. Metal-semiconductor junctions such as graphene/TMD heterostructures give rise to Schottky barrier height (SBH), whereas junctions between two semiconductors, such as TMD/TMD heterostructures, are categorized by their band offsets. Based on the band offsets, semiconductor-semiconductor interfaces can be either straddling gap (type-I alignment) or staggered-gap type (type-II alignment). MoSe$_2$/ WS$_2$, MoTe$_2$/ WSe$_2$, and WSe$_2$/ WTe$_2$ are examples of the former heterojunction-type, whereas MoS$_2$/ WS$_2$ and MoSe$_2$/ WSe$_2$ are the examples of the latter category~\cite{PantNanoscale16}. 

\subsection{Band alignment}
\begin{figure}[ht!]
	\centering
	\includegraphics[width=\columnwidth]{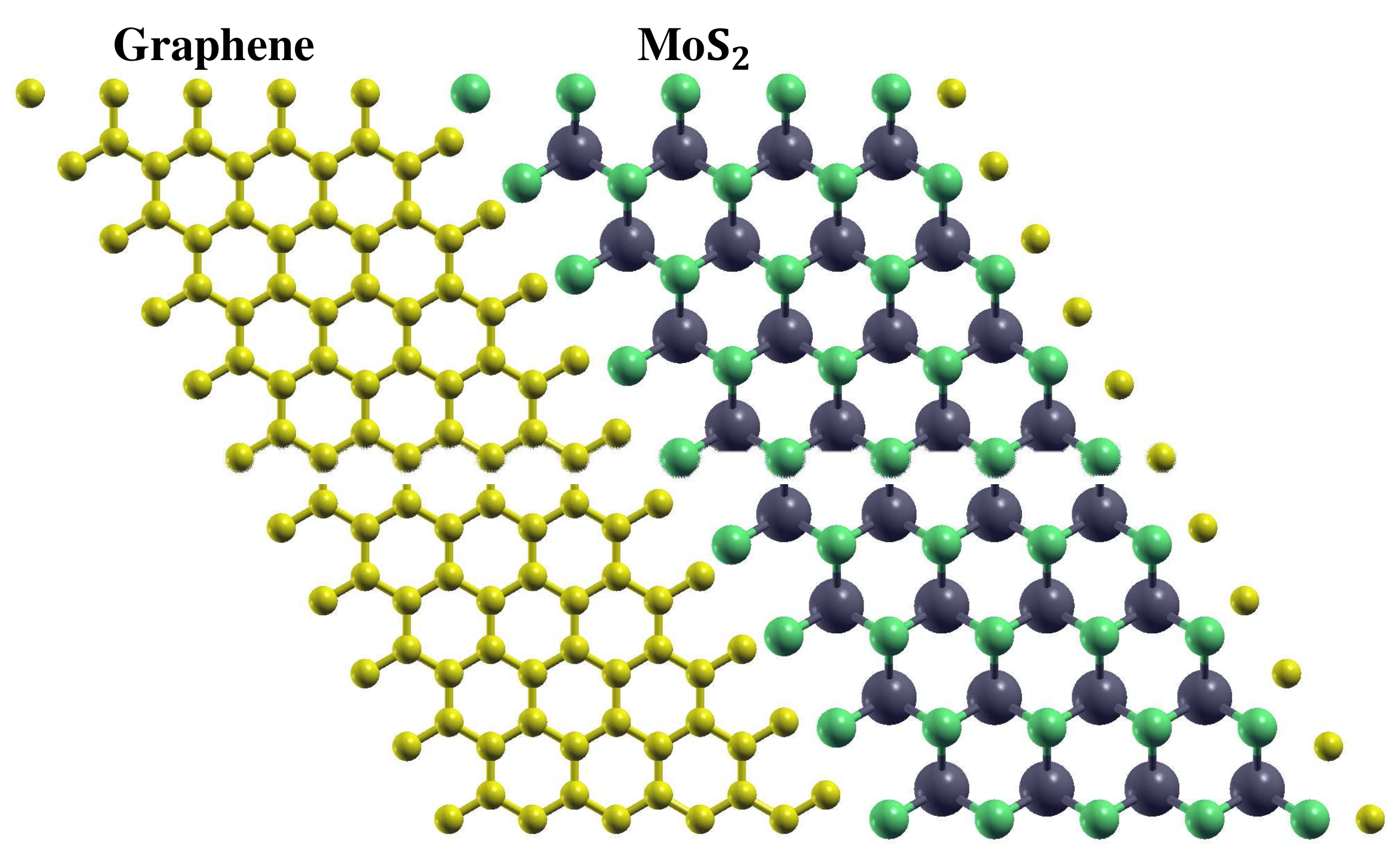}
	\caption{shows the schematic of a graphene/MoS$_2$ interface drawn in Quantum Espresso. The yellow, blue and green atoms represent carbon, molybdenum and sulphur atoms, respectively.}
	\label{fig:figure3e}
	\label{overflow}
\end{figure}
\begin{figure}[ht!]
	\centering
	\includegraphics[width=\columnwidth]{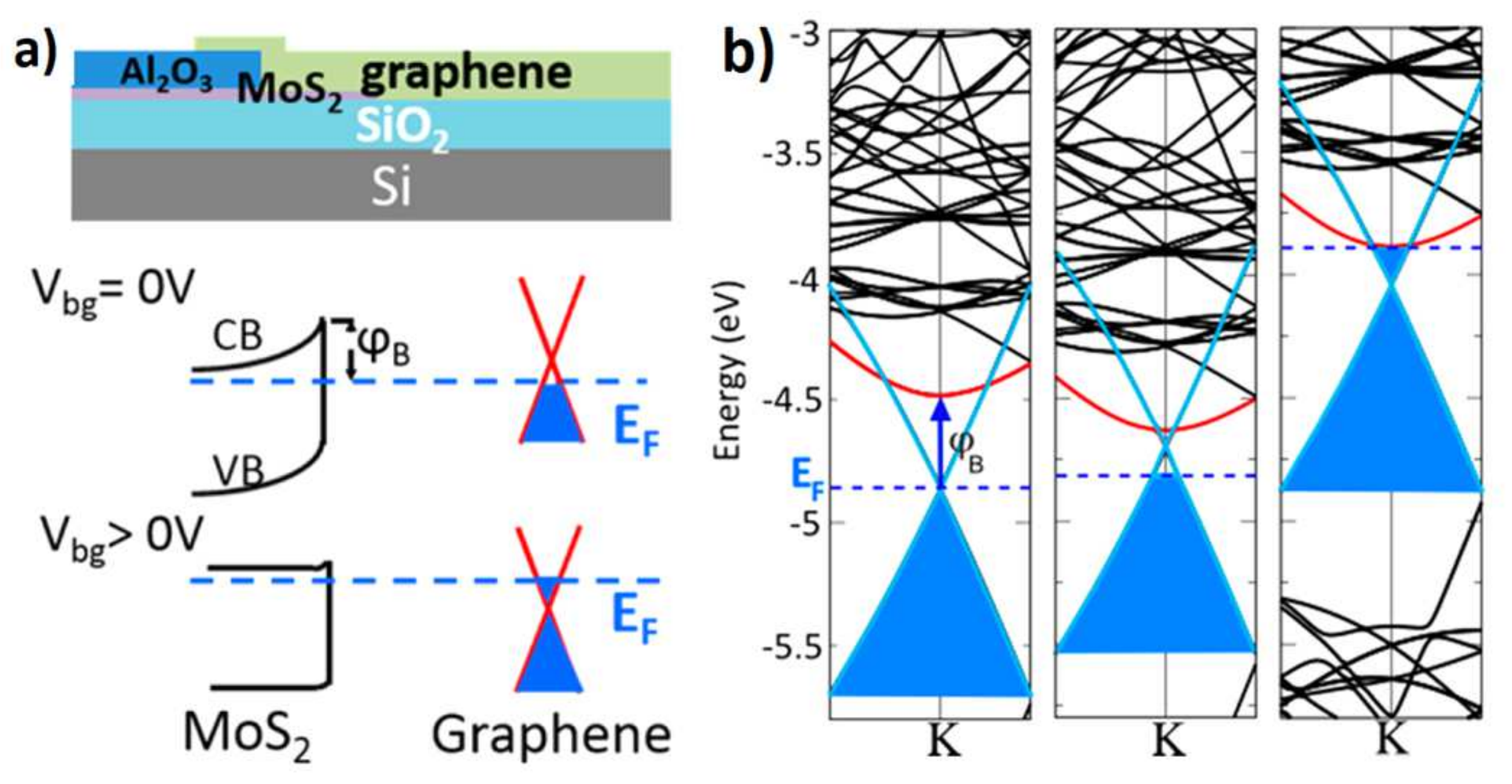}
	\caption{\textbf{(a)} shows the schematic of the experimental set-up and band diagram for unbiased and biased gate conditions respectively. \textbf{(b)} Calculated bandstructures of the graphene/MoS$_2$ interface at zero bias and doping (left), zero bias with finite doping concentration of 11.05$\times$ 10$^{12}$ cm$^{2}$ (middle) and 80 V at the same doping (right). Reprinted
    with permission from \cite{YuNanoLett14}. Copyright (2014) American Chemical Society.}
	\label{fig:figure3d}
	\label{overflow}
\end{figure}

 
There are different approaches to computing as well as aligning the electronic structure at the junction. One such approach is atomistic, based on first principles described by the widely-used Density Functional theory (DFT). In practice, most popular DFT implementations use a periodic basis, which makes it challenging to simulate non-periodic structures such as interfaces. A commonly used work-around is to make the structure artificially periodic within a larger expanded unit cell, called supercell shown in Figure~\ref{fig:figure3e}. One of the major drawbacks of this method is that it requires large supercells to effectively isolate the interfaces in adjacent supercells from each other, thus magnifying the computational cost. Therefore, simpler and more efficient methods, validated by DFT calculations, are desirable. The most widely used such empirical model is the Schottky-Mott rule~\cite{YuNanoLett14,MajeeSciRep17}, which says that the vacuum energy levels of the semiconductors on either side of the junction must be aligned, and thus allows band bending at the interface to account for different work functions and electron affinities. However, Schottky-Mott rule does not capture the charge redistribution that might happen when two materials are brought in contact. In an interface comprising of two three-dimensional materials, a local charge transfer leads to the formation of a plate capacitor. In contrast to this plate capacitor, a 2D heterojunction forms a pair of line dipoles~\cite{YuNanoLett16}. In 2D heterojunctions, it has been shown that the effect of this interface dipole becomes negligible when the overall device dimensions are much larger than the characteristic junction-width, typically about 10 nm~\cite{Zhang2DMater17}. Consequently, the band alignment in such heterojunctions was found to be less sensitive to the details of interfacial structure. Furthermore, the alignment was found to closely follow the Schottky-Mott rule~\cite{MajeeSciRep17} that is widely used to describe 3D interfaces. Figure~\ref{fig:figure3d}(a) shows an schematic of band-bending at the interface for unbiased (V$_{bg}$=0) and biased (V$_{bg}$>0) gate conditions. The bandstructure of graphene with respect to MoS$_2$ at the interface is shown in Figure~\ref{fig:figure3d}(b) for different biasing conditions. Besides intrinsic band-bending, there could be additional band-bending due to extrinsic factors such as local strain arising from lattice mismatch~\cite{WeiPhysChemChemPhys17}, confinement effects~\cite{MuMaterResearchExpress18}, and the presence of impurities~\cite{LyNatCommun16} at the junction. All these additional effects could change the transmission coefficient of carriers across the barrier, which in turn, will affect GB resistivity (see Equations~\ref{eq:EnergyTransmission} and ~\ref{eq:RGB}).

\subsection{Transport in graphene/MoS$_2$ heterojunctions}
Earlier studies used metals for contacts in MoS$_2$ FETs, whose performance was largely undermined by the contact resistance at their interface. In general, it is difficult to achieve zero energy barrier between a semiconductor and metal, which is an important criterion that influences device performance. Since graphene is a semi-metal whose work function can be tuned, it was found to form Ohmic contact with MoS$_2$~\cite{YuNanoLett14, LogotetaSciRep14}. This led researchers to investigate electrical transport in lateral heterojunctions with a particular focus on graphene/MoS$_2$ heterostructures~\cite{YuNanoLett16}. The interface resistance between graphene and MoS$_2$ has been quantified as a function of gate voltage and found to vary with the height of the potential barrier between them, which was found to depend on the band alignment and vary with carrier concentration~\cite{BehranginiaSmall17}.  

\begin{figure*}[ht!]
	\centering
	\includegraphics[width=6.5in]{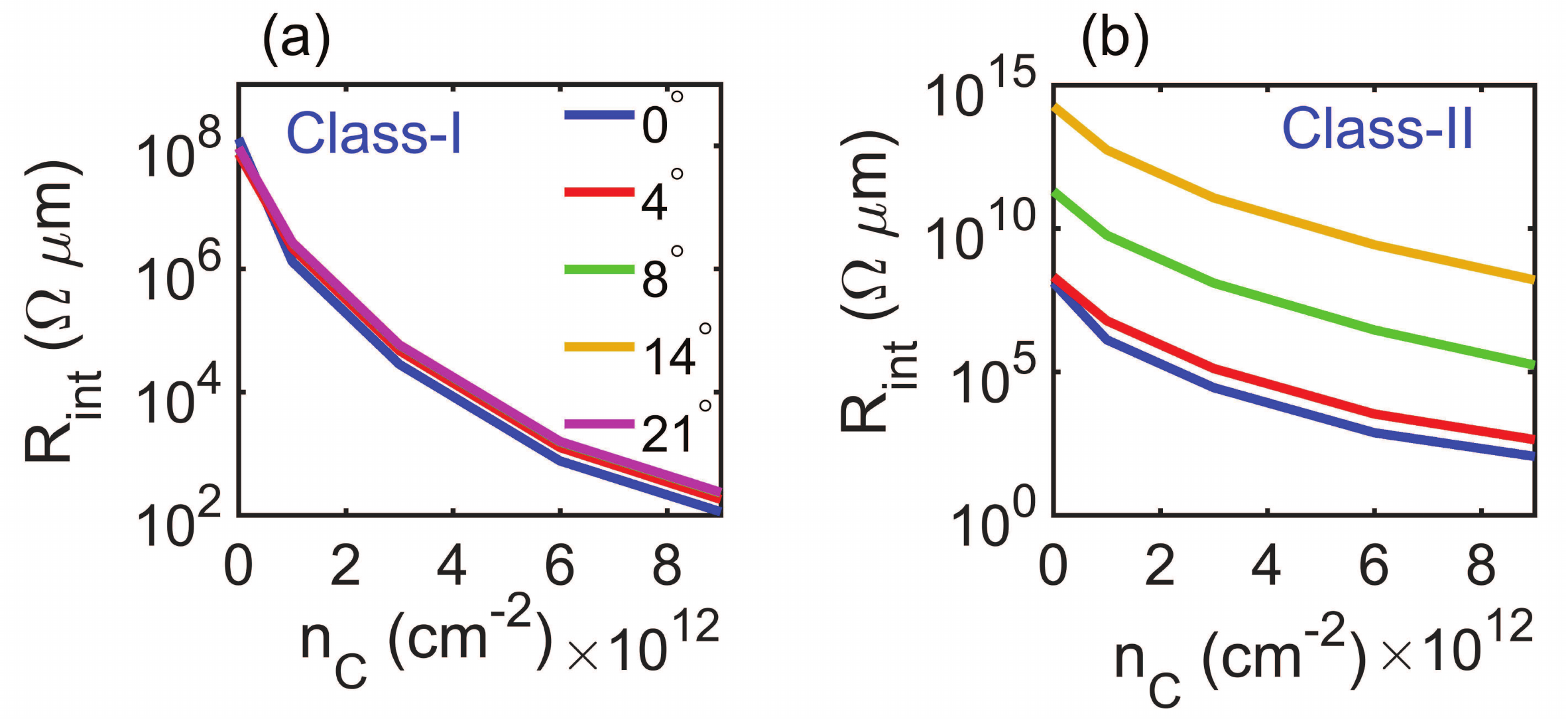}
	\caption{Interface resistivities vs. carrier concentration for various misorientation angles in graphene/MoS$_2$ \textbf{(a)} Class-I and \textbf{(b)} Class-II heterojunctions. These figures are taken from [\citenum{MajeeSciRep17}].}
	\label{fig:figure4}
	\label{overflow}
\end{figure*}

Despite several studies, both theoretical~\cite{LiuJPhysChemLett15} as well as experimental~\cite{TianSciRep14}, there is little known about the impact of misorientation angle on graphene/MoS$_2$ lateral heterojunctions. Sun et al.~\cite{SunCompMaterSci17} carried out first-principles calculations to study the relation between contact geometries and electrical properties of graphene/MoS$_2$ heterostructures in four configurations---zig-zag/zig-zag, armchair/armachair, armchair/zig-zag, and zig-zag/armchair. They found that MoS$_2$ is metalized due to the occurrence of band gap states at the contact interface which also reduces the Schottky barrier height (SBH) in all the four configurations. Recently, we systematically investigated the effect of misorientation angles on interface resistivity at graphene/MoS$_2$ interface~\cite{MajeeSciRep17} and found that the interface resistivity in symmetric graphene/MoS$_2$ interfaces (Class-I heterojunctions) is less sensitive to mismatch angle as shown in Figure \ref{fig:figure4}(a), but the resistivity is much larger than those of symmetric homojunctions. On the other hand, Figure \ref{fig:figure4}(b) shows that the resistivity of asymmetric graphene/MoS$_2$ interface (Class-II heterojunctions) strongly depends on mismatch angle. Like in graphene tilt GBs, this is also due to the underlap between CBs of graphene and MoS$_2$ on rotation of their BZs.

\subsection{Resonant tunneling diodes in 2D lateral heterostructures}

Tuning the band alignment at lateral heterojunction interface can open up interesting applications by optimizing their electronic and optical properties. Using one-step tube-moving chemeical vapor deposition method~\cite{PanJAmChemSoc18}, a lateral WS$_2$/WS$_{2(1-x)}$Se$_{2x}$ hetrostructure was synthesized. Controlling the ratio of WSe$_2$ and WS$_2$ in evaporaton sources enables the tuning of the band alignment in heterostructures. Another approach to tuning the bandgap~\cite{KomsaPRB13} in MoS$_2$ nanoribbons is introducing the sulfur vacancies, which enabled the creation of double barrier quantum well structure (DBQW). Following this approach~\cite{AkhoundiJCE17} and using NEGF based on tight-binding calculations, a DBQW of A-MoS$_2$ nanoribbons with two sulphur line vacancies was simulated. The resultant analysis showed that the double barrier structure has a PVR (peak-and valley-current ratio) of about 78 at room temperature.

\begin{figure}[ht!]
	\centering
	\includegraphics[width=\columnwidth]{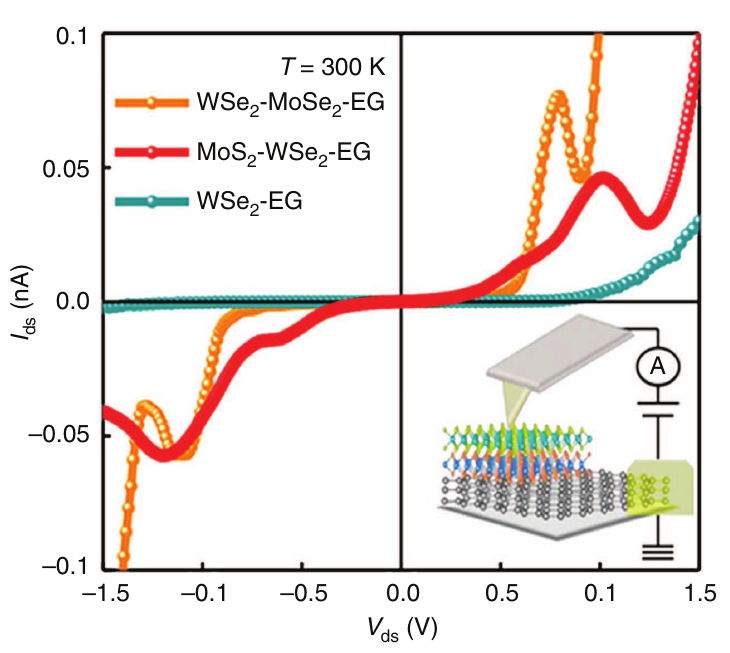}
	\caption{Resonant tunnelling and negative differential resistance (NDR) in heterostructures created by vertically integrating graphene with the monolayer transition dichalcogenides. Experimental current-voltage characteristics for different combination of dichalcogenide–graphene interfaces demonstrating spectrally narrow room temperature NDR. The schematic of the experimental set-up for the measurement in this system showed in inset. Reprinted with permission from \cite{LinNcomm2015}. Copyright (2015) Author(s), licensed under the Creative Commons Attribution 4.0 License.}
	\label{fig:figure5}
\end{figure}

One of the first demonstrations~\cite{LinNcomm2015} created pristine multi-junction heterostructures by direct synthesis based on graphene, MoS$_2$, MoSe$_2$ and WSe$_2$. This lead to the creation of resonant tunneling diode (RTD) in an atomically-thick stack with negative differential resistance (NDR). The spectrally-narrow NDR created in this structure, as shown in Figure \ref{fig:figure5}, is superior to manually stacked heterostructures that were studied previously. TMDs in the symFET architecture ~\cite{CampbellACS15} have been shown to achieve PVR up to $10^9$ compared to few hundred observed in graphene or III-V RTDs. Also, the ability to achieve such high PVR even in nanoscale devices makes them promising candidates for digital logic applications. A vertical heterostructure based on phospherene/rhenium disulfide (BP/ReS$_2$)  demonstrated NDR with a PVR up to 6.9 at 180 K ~\cite{ShimNatCom16}. A ternary inverter was also created from this heterostructure that exhibited three distinct logic values, demonstrating a proof of concept for future multi-valued logic devices.

\section{Vertical 2D heterostructures} 
For several device applications, the 2D material has to be placed on a substrate, typically SiO$_2$. Impurity scattering from the substrate significantly deteriorates carrier mobility as compared to their intrinsic phonon-limited mobility~\cite{CastroNetoRevModPhys09,ChenNatNano08,MartinNatPhys07}. To minimize the reduction of mobility of graphene on highly disordered SiO$_2$~\cite{IshigamiNanoLett07,HuangPRL07}, heterojunction made up of graphene and hexagonal-boron nitride had been proposed which demonstrated improved carrier mobility~\cite{DeanNatNano10}. Moreover, the possibility of fabricating low power devices using band-to-band tunneling (BTBT) and achieving steep subthreshold slope below 60 mV/dec paved the way for devices made up of vertical heterostructures. Vertical heterostructures, which are typically referred to as \emph{van der Waals} heterostructures, have been recently found to be of research interest because of the large contact area at the interface between the two materials and absence of dangling bonds. As the name suggests, the layers are held together in the vertical direction with weak van der Waals forces as compared to the strong covalent bonding in lateral heterojunctions. Exciting physics and potential device applications in stretchable/flexible electronics have been demonstrated by making field-effect transistors~\cite{BritnelScience12,GeorgiouNatureNano12}, memory devices ~\cite{HongACSnano11,SupChoiNatCommun13,BertolazziACSnano13}, and photodetectors~\cite{ChengNanoLett14,DengACSnano14} by vertically stacking two-dimensional materials. 

Earlier studies used mechanical exfoliation to peel off layers of various materials from their bulk counterparts and then used a micro-manipulator to stack and align the layers in a vertical configuration. The mechanical transfer techniques make the interfaces between layers highly susceptible to impurities and contaminants~\cite{HaighNatMater12,YangNatMater13}, which made it difficult to fabricate vertical heterostructures with sharp and clean interfaces. In subsequent years, researchers devised novel fabrication techniques by employing direct-synthesis of such vertical stacks using single-step vapor-phase growth process, which yields much cleaner and sharper interfaces~\cite{GongNatMater14,LinACSnano14}. Recently, CVD growth of centimeter-scale 2D vertical heterostructure has also been demonstrated~\cite{IslamNanoLett17}. All the aforementioned fabrication techniques and potential applications of 2D van der Waals heterostructures have been discussed in great details in the following reviews~\cite{LimChemMater14,LotschAnnRevMaterResearch15}.

\subsection{Charge transport in vertical 2D heterojunctions}\label{sec:vertical}
The charge conduction across different materials in a vertical stack takes places either through BTBT or modulation of SBH at a 2D-3D interface. BTBT, also called interlayer tunneling, was first demonstrated in 2D vertical heterostructures by Britnell et al.~\cite{BritnelScience12} where an ultrathin layer of hexagonal boron nitride (hBN) was sandwiched between two graphene monolayers. The tunneling current was regulated by varying the gate voltage, which altered both carrier densities in graphene layers as well as internal barrier voltage. Georgiou et al.~\cite{GeorgiouNatureNano12} replaced hBN with WSe$_2$ to improve the ON-current and switching ratio from $10^5$ to $10^6$. Roy et al.~\cite{RoyACSNano15} fabricated a MoS$_2$/WSe$_2$ vertical heterostructure  in a dual-gate architecture, where the carrier densities of each material were modulated independently using a pair of symmetric gates. The same device exhibited large reverse-biased current caused by BTBT, an Esaki diode behavior with negative differential resistance (NDR) or a forward-biased rectifying behavior depending on different gating conditions. Nourbakhsh et al.~\cite{NourbakhshNanoLett16} attributed the presence of NDR in such devices to lateral BTBT. Their results also confirmed that lateral BTBT is more dominant than the tunneling through the overlapped vertical section.

Recently, Sarkar et al.~\cite{SarkarNature15} demonstrated a 2D/3D semiconductor planar transistor based on BTBT which achieved the record-low subthreshold swing with a minimum value of 3.9 mV/dec and an average of 31.1 mV/dec (5.5, 12.8, 22 mV/dec) over four (one/two/three, respectively) decades of drain current. They named it ATLAS, which stands for atomically thin and layered semiconducting-channel tunnel FET. They strategically chose germanium as the 3D material among all other Group IV and III-V compounds and MoS$_2$ as the 2D material among other TMDs to form a staggered heterojunction with a low tunneling barrier height. Another interesting vertical three-terminal device is called \emph{barristor}, a portmanteau of barrier and transistor, in which the gate voltage is used to regulate tunneling through the SBH formed at their interface instead of BTBT. Yang et al.~\cite{YangScience12} demonstrated the first graphene barristor using a tri-layer stack (metal/graphene/silicon) with a large on-off ratio of $10^5$. Yu et al.~\cite{YuNatMater12} used the same idea to fabricate a 2D-2D barristor using metal/MoS$_2$/graphene stack to improve the current density by about two to five orders of magnitude while maintaining a high on-off ratio ($>10^3$). Recently, twist angle (misorientation angle between layers in vertical direction) has been used in bilayer graphene to demonstrate superconductivity at few \emph{magic angles}, the first being at $1.1^\circ$. At such magic angles, twisted bilayer-graphene exhibits flat bands near the Dirac point resulting in increase in density of states~\cite{CaoNature18}. In another recent work, Liao et al.\cite{LiaoNatcommun18} found that twist angles play a crucial role in determining vertical conductivity in a graphene/MoS$_2$ stack. Like our recent study on lateral graphene/MoS$_2$ heterojunction\cite{MajeeSciRep17}, the conductivity has been found to be highest for twist angle equal to $0^\circ$ and minimum for $30^\circ$ in graphene/MoS$_2$ vertical heterojunction. The reason of this has been attributed to the reduction in transmission coefficient with twist angles.      

\subsection{Electronic thermopower in 2D heterostructures}

In addition to having large electron mobilities, some 2D materials have exhibited interesting thermoelectric (TE) properties, including possessing a large Seebeck coefficient, also called thermopower. It is defined as the ratio of the voltage produced by thermally diffusing electrons and the temperature gradient that is driving their diffusion from the hot side to the cold. Research in this area started with the measurements of TE transport in graphene \cite{ZuevPRL09}, which exhibited anomalous TE transport of Dirac particles \cite{WeiPRL09}. More recently, the enhanced TE Seebeck coefficient in graphene \cite{GhahariPRL16} was attributed to the role of hydrodynamic transport through inelastic scattering, which also leads to a violation of the Mott relation for the Seebeck coefficient (not to be confused with the Schottky-Mott rule for band alignment)
\begin{equation}
\left. S_{Mott} = -\frac{\pi^2 k_B^2 T}{3|e|\sigma}\frac{d \sigma(E)}{d E}\right\vert_{E=E_F}.
\end{equation}
\noindent Proposals to enhance the TE properties of graphene used heterostructures \cite{DragomanAPL07} and functionalization \cite{KimNL15}. Graphene also exhibited a significant phonon drag component of the Seebeck coefficient \cite{KoniakhinEPL13}, which is defined as the additional thermopower caused by the exchange of momentum between the heat-carrying lattice vibrations (phonons) and the electrons. This effect is particularly prominent at low temperatures where the diffusion thermopower is typically low.

Using both carrier and phononic engineering on a ZrSe$_2$/HfSe$_2$ single-layer superlattice~\cite{ZhangNanoS18}, a ZT of 5.3 in \textit{n}-type and 3.2 in \textit{p}-type device was achieved. First-principles calculations along with Boltzmann transport equation showed that this is achieved due to the high degenerate nature of the conduction bands in \textit{n}-type device. A partially overlapped graphene/graphene vertical heterostruture~\cite{DollfusAPL14} studied using atomistic tight-binding Hamiltonian demonstrated that the ZT can reach unity at room temperature. An analysis~\cite{LiangSciRep17} on van der waals heterostructures with multilayer TMDs sandwiched between two graphene electrodes identified WSe$_2$ and MoSe$_2$ as the ideal TMDs that can provide high TE conversion efficiency.  Graphene/\textit{h}-BN/graphene vertical heterostructures are synthesized~\cite{ChenNano2015} to study their TE performance. Measurements uncovered a significant Seebeck coefficient at the material interfaces, which makes such heterostrutures suitable candidates for TE applications. Twisted bilayer graphene vertical heterostructures are studied extensively due to their interesting properties like magic angles that can induce superconductivity, described in the previous section. With regards to their thermoelectric properties, measurements on twisted graphene bilayers~\cite{GhoshNanoLett17} showed exceptional cross-plane thermopower, attributed to phonon drag from out-of-plane phonon modes (ZA/ZO'). Besides twisted graphene bilayers, MoS$_2$ has been shown to also exhibit a phonon drag effect \cite{BhargaviJPCM14}.
   
\section{3D heterostructures and superlattices}\label{sec:3Dhetero}

With the discovery of NDR in narrow highly doped germanium \textit{p}-\textit{n} junctions by Esaki~\cite{EsakiPRB1958}, a new era of electronics based on quantum tunneling devices was ushered in. These \textit{p}-\textit{n} junctions were called tunnel diodes as they demonstrated interband tunneling of carriers from the valence band on the \textit{p}-side to conduction band on the \textit{n}-side and vice versa. For this discovery, Leo Esaki shared the 1973 Nobel Prize in Physics with Brian Josephson. Around the same time, Tsu and Esaki~\cite{EsakiTsuIBM1970} found that in the direction of a one-dimensional periodic potential, such as that found in a heterostructure (superlattice), electrons are localized to discrete energy states analogous to electrons in a two-dimensional electron gas. Further, due to the comparable dimensions of these potentials with electron wavelengths, the wave nature of electron leads to phenomena such as interference and tunneling. In such structures, when the energy of electrons coincides with one of the discrete energy states achieved by tuning the applied bias, electron can tunnel through the barriers and can have near-perfect transmission. With further increase in the applied bias, the amount of available electrons decreases, thus reducing the current flowing through the structure. Ultimately, the current increases again when high enough bias is applied to cause thermionic emission of electrons over the top of the potential barrier. This distinct feature is popularly known as \emph{negative differential resistance} and has found numerous applications in devices such as high-frequency oscillators, frequency converters, and detectors and also exhibits great potential in high-speed logic devices and switches.

\subsection{Dual barrier heterostructure: resonant tunneling diode} 

Shortly after the pioneering discoveries of Tsu and Esaki, it was shown~\cite{TsuEsaki73} that the same effect can be achieved using a finite superlattice. With the advancements made in fabrication process, a finite superlattice with a double barrier is created~\cite{ChangEsakiAPL74} in GaAs by sandwiching it with Ga$_{1-x}$As$_x$ to form the potential barriers and demonstrated NDR phenomenon. A basic RTD is made up of a double-barrier heterostructure with nanoscale barriers made from a heavily doped semiconductors. Beside tunneling, there are other physical processes like interactions of carriers with lattice, impurities, surface roughness, other carriers and alloy disorder. A realistic physical model of RTDs including all these processes requires advanced quantum transport theory like density matrix, Wigner functions, and non-equilibrium Green's functions (NEGF).

\subsection{Progress in numerical simulation of 3D heterostructures}
The semi-classical Boltzmann transport equation, which is used widely to simulate electron and thermal transport in electronic devices, is not adequate to capture quantum effects that are predominant in RTDs. Various quantum-transport frameworks such as the Wigner formalism and the related density matrix approach, as well as non-equilibrium Green's functions (NEGF), have been employed to study transport in RTDs. NEGF was one of the first methods used to study the transport in RTDs. Lake and Datta \cite{LakeDuttaPRB92} studied the effect of energy broadening and inelastic scattering and reported an enhanced valley current due to inelastic scattering with simultaneous enhancement of occupation of the resonant states. Time-dependent transport capabilities of NEGF were demonstrated by simulating RTD~\cite{JauhoPRB94} in mesoscopic region. Nam Do et al.~\cite{ButtikerELett96} used fully self-consistent non-equilibrium Green's function approach to study the the impact of quantum-well width, the barrier thickness and the temperature, and showed their effect on PVR. PVR, as the name implies, is the ratio between the currents at the peak before the onset of the NDR and the valley after it, which helps to determine the feasibility of using the RTD for device applications.

Wigner formalism is one of the first methods~\cite{FrensleyPRB87} used to simulate the transport behavior of quantum-well RTD and their NDR. Biegel and Plummer~\cite{BiegelPRB96} implemented a self-consistent Wigner-function-based quantum device simulation using the RTD as a test case. A comparison between different iterative methods was presented here to solve the Wigner function. It concluded that transient Gummel approach is reasonably accurate with low computational resource requirement for device simulations. A unified approach~\cite{NedjalkovPRB04} was proposed to merge Wigner functions with semi-classical Boltzmann transport equation and treat the scattering term as a generating term. In that study, interference effects due to the Wigner potential were associated with particle generation having statistical weights.

\subsection {Thermoelectric properties of superlattices}\label{sec:TEprops}

Quantum effects in heterostructures have a large impact on their thermoelectric properties, affecting both the Seebeck coefficient $S$, which is related to the average transport energy per carrier, and the conductivity $\sigma$. Together, the Seebeck coefficient and conductivity constitute the TE power factor $PF=S^2 \sigma$, which is central to the TE conversion efficiency captured by the figure-of-merit $ZT=S^2 \sigma/\kappa$. $\kappa$ is the sum of electronic and lattice contributions to the thermal conductivity. In their seminal work, Hicks and Dresselhaus\cite{HicksPRB93QWell} predicted a one order of magnitude increase in figure-of-merit (ZT) for quantum well superlattices made from Bi$_{2}$Te$_{3}$. High anisotropy along certain direction at low widths of quantum wells are shown to be the reason for this enhancement. This kicked off extensive research into understanding their thermoelectric properties and possible use in thermoelectric generators. Their predictions were later revised~\cite{SofoMahanAPS94} when it was shown that the enhancement is controlled by period of the superlattice and not by the width of the well. Also, the overall $ZT$ is projected to be less than one when tunneling is introduced between quantum wells. This idea was experimentally validated using PbTe/Pb$_{1-x}$Eu$_x$Te superlattices~\cite{Hicks1996} with a five fold increase in $ZT$ compared to bulk PbTe. A more extensive analysis~\cite{Lin-Chung1995,BroidoAPS97} for the case which includes a non-zero barrier and effects of carrier tunneling, revealed that the in-plane thermoelectric performance is not superior to bulk systems. Later short-period Si/Ge superlattice structures were shown~\cite{ColpittsICT98} to provide higher $ZT$ compared to thin-film SiGe and bulk Si-Ge alloys, which was attributed to the reduction in thermal conductivity. A further improvement~\cite{DesselhausAPS99} in $ZT$ is reported in Si/Ge superlattices by applying strain to further tune the conduction band structure. Experimental studies demonstrated the thin-film thermoelectric coolers using both single heterostructure and superlattice structures from SiGe/Si~\cite{ShakouriAPS99,ZengElett1999,FanAPL01,FanElett2001,BowersJAP01,BowersOptE00}.

Using an exact numerical solution, a complete treatment of powerfactor in PbTe quantum well and quantum wire superlattices showed~\cite{BroidoAPlett00} weaker dependencies on potential barrier. Extending the study to GaAs quantum wire superlattices~\cite{BroidoPRB01}, it was established that significant enhancements in power factor can be achieved from restricting phonon transport.  PbTe/PbTeSe-based superlattices are experimentally shown~\cite{RamaAPL05,BauerAPL02} to improve $ZT$ by 50\% relative to bulk PbTe. This can be attributed to a reduction in thermal conductivity, an affect of the increase in scattering due to alloying. Later, majority of studies used this approach to improve $ZT$. Ge quantum dot in Si quantum dot superlattices were shown~\cite{ChenNanoT00} to improve $ZT$ by reducing the thermal conductivity due to scattering of phonons. Theoretical calculations of TE properties in superlattice nanowires~\cite{LinPRB03} based on lead salts (PbS, PbSe and PbTe) exhibited a significantly higher ZT values, especially for 5 nm diameter wires. 

\begin{figure}[ht!]
	\centering
	\includegraphics[width=\columnwidth]{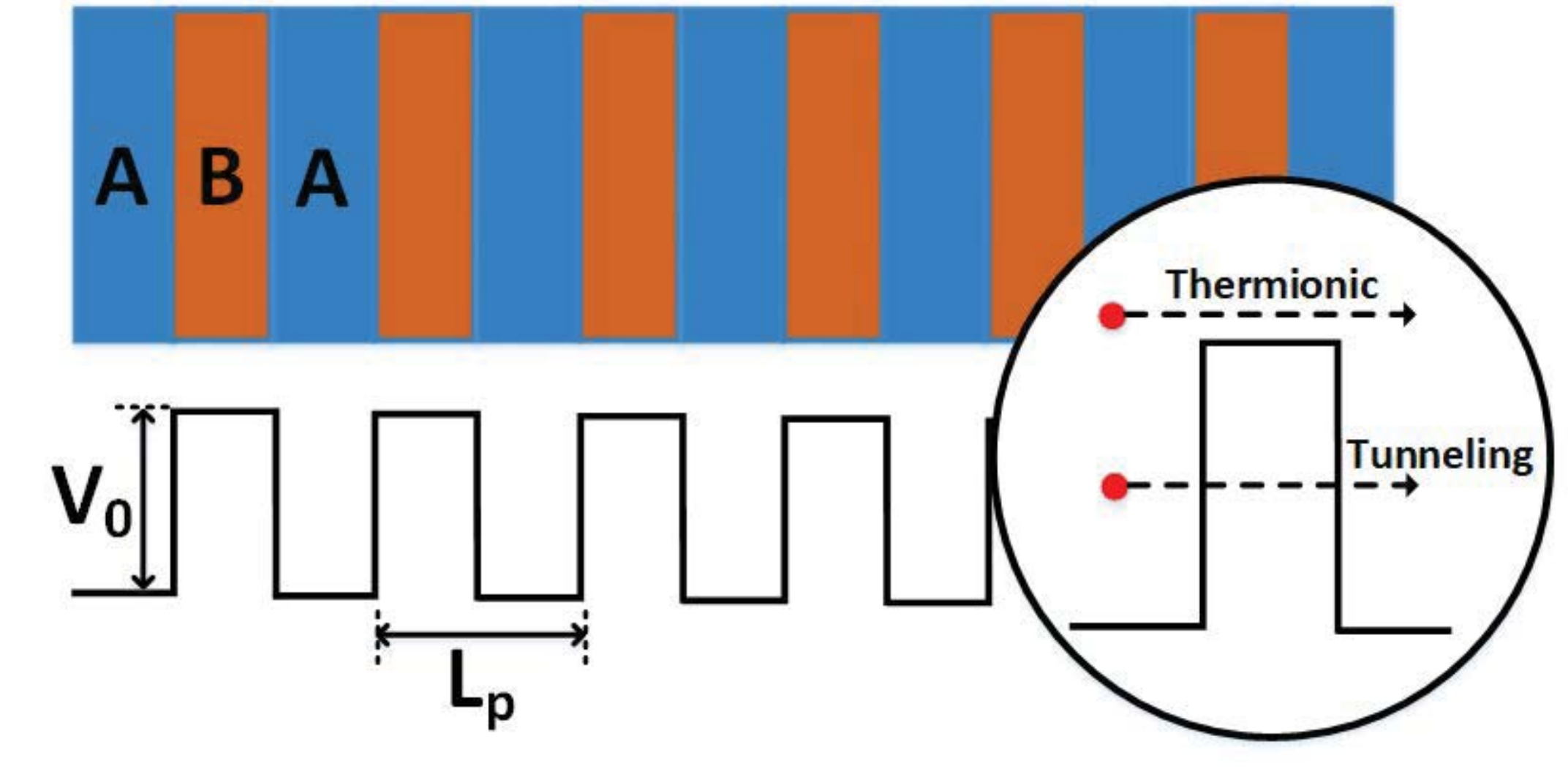}
	\caption{Periodic potential barriers can be used to understand the superlattice or heterostructure. Here, V$_0$ is the height of potential barrier and L$_p$ is the period of the potential barriers. Due the presence of potential barriers, carrier transport in the system is controlled by thermionic emission over the barrier and tunneling through the barrier. Adapted from \cite{KomminiJPCM2018}}
	\label{Gdevice}
	\label{fig0}
\end{figure}

Even though significant improvement in $ZT$ is reported by restricting thermal conductivity, further improvement can be achieved by tuning the TE power factor. Many studies proposed mechanisms and ways to improve power factor in superlattices by tuning the band structure~\cite{BalandinAPL03,BianPRB07,HarmanJEM00}. Energy filtering, a process of restricting movement of carriers with kinetic energy smaller than barrier height, has been consistently found to enhance the overall power factor. It raises the average energy of carriers and thereby, its Seebeck coefficient. Nanocomposites~\cite{KimJAP11,ChenNanolett11,ZideJAP10}, superlattices~\cite{YokomizoAPL13} as well as single and multiple barrier structures have been used to introduce some form of energy filtering in a material. Studies~\cite{Neophytou2013,Neophytou2016,DragomanAPL07,ChengPhyE2012} tried to understand the effect of potential barrier smoothness and their structure on thermoelectric performance. Using a series of gates placed periodically in graphene, creating periodic potentials \cite{DragomanAPL07,ChengPhyE2012} that resulted in a drastic increase in Seebeck coefficient. Thermionic emission over the barrier and tunneling through the barrier, control transport in these structures as shown in Figure~\ref{Gdevice}. Semi-classical transport (diffusive transport) theory like Boltzmann transport equation is capable of recreating the thermionic emission over potential barriers. Quantum effects like tunneling are simulated using density matrix formulation, Wigner formalism and  non-equilibrium Green's functions.  Here we aim to understand the strengths and limitations of these approaches in simulating the quantum effects along with the diffusive transport.

\subsubsection {Wigner formalism}
In the semi-classical Boltzmann transport equation (BTE), widely used in device simulation, electrons are treated as point particles~\cite{JacoboniRMP83,MohamedTED14}. Despite its widespread use, this approach is unable to explain some device effects, such as underestimating the threshold voltage in ultra-thin body MOSFET~\cite{Chin97IEEE,Wu96IEEE}, and the carrier interactions with the rapid potential changes across heterojunctions. The impact of sharp spatially-varying potentials on transport in a superlattice can be simulated using the Wigner formalism. The Wigner formalism~\cite{Nedjalkov11,Wigner32APS} includes an additional quantum evolution term in addition to drift term that can capture the spatial variation in potential. The Wigner formalism has been the subject of an extensive recent review article~\cite{WeinbubAPR18}. BTE can be modified as follows to include Wigner formalism for rapid varying spatial potentials:
\begin{equation}\label{nSWBTE}
\left(\frac{\partial }{\partial t}+v_r{\nabla }_{r}+\frac{eF}{\hslash}{\nabla }_{k}\right){f}_{w }\left(r,k,t\right)=Q{f}_{w }\left(r,k,t\right)+{\left(\frac{\partial f_{w}}{\partial t}\right)}_{coll}
\end{equation}
where ${f}_{w }\left(r,k,t\right)$ is called Wigner distribution function, which is written as 
\begin{equation}
{f}_{w }\left(r,k,t\right)=\frac{1}{{2\pi}}\int d{r}^{\prime }{e}^{-i{r}^{\prime }k}\rho \left(r+\frac{{r}^{\prime }}{2}, r-\frac{{r}^{\prime }}{2}\right)
\end{equation}
where the mixed state is represented by the density operator $\rho$. Here the coordinates $r$ and ${r}^{\prime}$ represent the center of mass and spread of the electron wave packet, respectively. The potential operator $Q{f}_{w }\left(r,k,t\right)$, often called the quantum evolution operator, is given as
\begin{equation}\label{qw}
Q{f}_{w }\left(r,k,t\right)=\int d{k}^{\prime }{V}_{w}\left(r,k-{k}^{\prime }\right){f}_{w }\left(r^{\prime},k,t\right)
\end{equation}
where the Wigner potential $V_w(r,k)$ is itself also obtained through a Wigner transform
\begin{equation}\label{vw}
{V}_{w}\left(r,k\right)=\frac{1}{i\hslash {\left(2\pi \right)}^{d}}\int d{r}^{\prime }{e}^{-i{r}^{\prime }k}\left(V\left(r+\frac{{r}^{\prime }}{2}\right)-V\left(r-\frac{{r}^{\prime }}{2}\right)\right)
\end{equation}
of the electrostatic potential $V$ across the material, expressed in the new Wigner coordinates $r$ and $r'$. Eq. \ref{vw} can be simplified as 
\begin{equation}\label{vwsim}
{V}_{w}\left(r,k\right)=\frac{2}{\pi \hslash}\> Im \{e^{2ikr} \hat{V}(2k)\}
\end{equation}
where $\hat{V}(k)$ is spatial Fourier transform of $V$ 
\begin{equation}
\hat{V}(k)=\underset{-\infty}{\overset{\infty }{\int}} V(r) e^{-ikr}dr .
\end{equation}
The potential operator $Q{f}_{w }\left(r,k,t\right)$ is usually decomposed into two components: a slowly-varying, often called classical, potential (such as the applied external bias) V$_{cl}$, and separate, rapidly-varying quantum-mechanical portion V$_{qm}$. Together, these two make up the potential according to 
\begin{equation}\label{clqm}
V(x)=V_{cl}(x)+V_{qm}(x).
\end{equation}
By including the additional effect of rapidly varying potentials in the BTE, the resulting steady-state Wigner-Boltzmann transport equation (WBTE) can be written as (from Eq. \ref{nSWBTE})
\begin{equation}\label{WBTE}
\left(v_r{\nabla }_{r}+\frac{eF}{\hslash}{\nabla }_{k}\right){f}_{w }\left(r,k,t\right)=Q{f}_{w }\left(r,k,t\right)+{\left(\frac{\partial f_{w}}{\partial t}\right)}_{coll}.
\end{equation}

The collision term helps to further incorporate the semi-classical effects like energy and momentum relaxation from time-dependent perturbation theory. The Wigner formalism permits us to couple the interdependencies between quantum and semi-classical effects, as it uses a phase-space formulation. This allows us to study the carrier energy relaxation in the presence of both phonon scattering and periodic potential barriers. The use of conventional semi-classical boundary conditions for contacts in the Wigner formalism has been found to result in unphysical results, especially in the coherent regime. This is due to non-unique solutions~\cite{RosatiPRB13,TajEPL06} and the nonlocal nature of Wigner approach. Dissipation and decoherence phenomena, even though they do lead to a unique solution, can also lead to unphysical results. Simplified local scattering models to define the dissipative transport can also cause negative probability-density~\cite{ZhanJCE2016,LottiPRB17}. Alternate approaches using Lindblad-type scattering superoperators ~\cite{RosatiPRB14,DolciniPRB13} and conditional wave functions~\cite{ColomesPRB17} have been proposed to solve these problems within the Wigner approach. These limitations are avoided in periodic systems, where contacts are replaced with periodic boundary conditions~\cite{KomminiJPCM2018}.

In our recent study~\cite{KomminiJPCM2018}, we developed a numerical model that combines a full bandstructure with the WBTE, with the goal of studying the impact of quantum effects on thermoelectric transport in semiconductors. It was implemented by solving the full-band WBTE iteratively in a manner analogous to the widely-used Rode's approach~\cite{RodePRB70,RodePRB71} to study the influence of shape, $V_0$, and $L_p$ on thermoelectric performance. In Rode's method, the accuracy of the solution to BTE is improved by including contributions from inelastic scattering processes, especially the in-scattering. This is achieved by splitting the BTE collision integral into two separate terms---the in-scattering and the out-scattering. Then the perturbation $g(k)$ to the equilibrium distribution function $f_0(k)$ (where $f_w(k) =f_0(k)+g(k)$) is solved iteratively until convergence. We refer the interested readers to Appendix A in \cite{KomminiJPCM2018} for a detailed derivation. The final iteration is expressed as
\begin{equation}\label{Rode_b}
\left. {g}_{i+1}(k)=\left({I(k)+\frac{eF}{\hslash }\frac{\partial {f}_{0}}{\partial k}-v(k)   \frac{\partial{f}}{\partial{r}}}\right) \middle/{{S}_{0}(k)} \right. .
\end{equation}
The iteration is started from an equilibrium distribution, such that the initial in-scattering term is zero and the first update is straightforward to compute. The first iteration corresponds exactly to the relaxation time approximation (RTA). Next, Rode's approach is applied to solving the WBTE by adding the Wigner potential contribution representing the additional forces arising from the potential variation in the structure. Then, the resultant perturbation to the distribution function in Eq. \ref{Rode_b} can be modified as
\begin{equation}\label{Wgi}
\left. {g}_{i+1}(k)=\left({I(k)+\frac{eF}{\hslash }\frac{\partial {f}_{0}}{\partial k}-v(k)   \frac{\partial{f}}{\partial{r}}+Qf_w}\right) \middle/{{S}_{0}(k)} \right..
\end{equation} 
To study the effect of periodic potential structure (shape and size), a square barrier with smoothening parameter $\beta$ and a smooth cosine shaped potential barrier are used. A generalized potential that is spatially periodic can be represented as
\begin{equation}
V(r)=\sum _{n=-\infty}^{\infty} V_p(r-nL_p) .
\end{equation}
The quantum evolution (Eq. \ref{qw}) (derived in Appendix B of \cite{KomminiJPCM2018}) of this generalized form for periodic potentials is 
\begin{equation}\label{fwgen}
Qf_w=\sum _{m=1}^{\infty} W_m(r) \left[f_w\left(r,k-\frac {m\pi}{L_p}\right)-f_w\left(r,k+\frac {m\pi}{L_p}\right)\right].
\end{equation}

The quantum evolution force for square barriers, 
\begin{equation}\label{gen_v}
V_p(r)=\frac {V_0}{2}\left\{ -erf[\beta(r-a)]+erf[\beta(r+a)] \right\}
\end{equation}
of height $V_0$ with smoothening factor $\beta$ and width $2a$ or $L_p/2$ is
\begin{equation}
W_m(r)=\frac {2V_0}{\pi \hbar m} e^{\frac {-m^2\pi^2}{\beta^2 {L_p}^2}} \sin\left(\frac {2\pi ma}{L_p} \right)\sin\left(\frac {2\pi mr}{L_p} \right).
\end{equation}

Similarly for a cosine shaped potential (applied in $r$ direction) of form
\begin{equation}\label{cos_poten}
V_q(r)={V_0} (1+\cos(K_{0}r))/2, \mbox{where} ~K_{0}= 2\pi/L_p
\end{equation}
the quantum evolution is obtained as (Appendix B in \cite{KomminiJPCM2018}),
\begin{equation}\label{fwcos}
Qf_w=W_m(r)\left[{f}_{w }\left(r,k-\frac{{K}_{0}}{2 }\right)-{f}_{w }\left(r,k+\frac{{K}_{0}}{2 }\right)\right]
\end{equation}
\begin{equation*}
\mbox{with quantum weight} \> W_m(r)=\frac{A \sin(K_{0}r)}{\pi \hslash}.
\end{equation*}
It should be noted that the quantum evolution force for a smooth potential is the first order approximation (m=1) to the quantum evolution force of a square barriers (Eq.~\ref{fwgen}).

\begin{figure*}[ht]
	\begin{subfigure}[b]{0.5\textwidth}
		\centering
		\includegraphics[width=\textwidth]{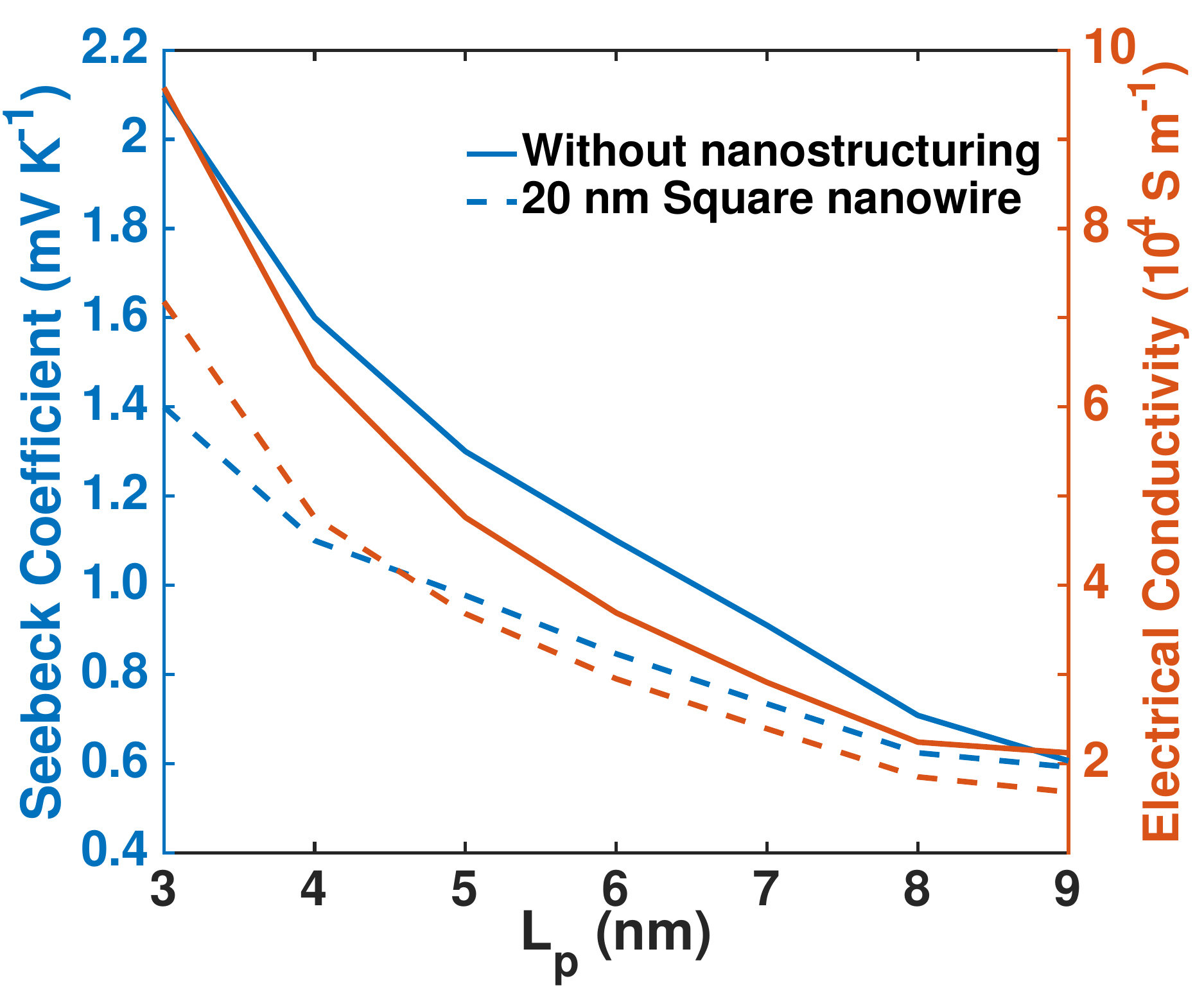}
		\caption{V$_0$ = 1.6 $k_B$ T and N$_D$ = 10$^{19}$ cm$^{-3}$}
		\label{3fig6_1}
	\end{subfigure}
	\begin{subfigure}[b]{0.5\textwidth}
		\centering
		\includegraphics[width=\textwidth]{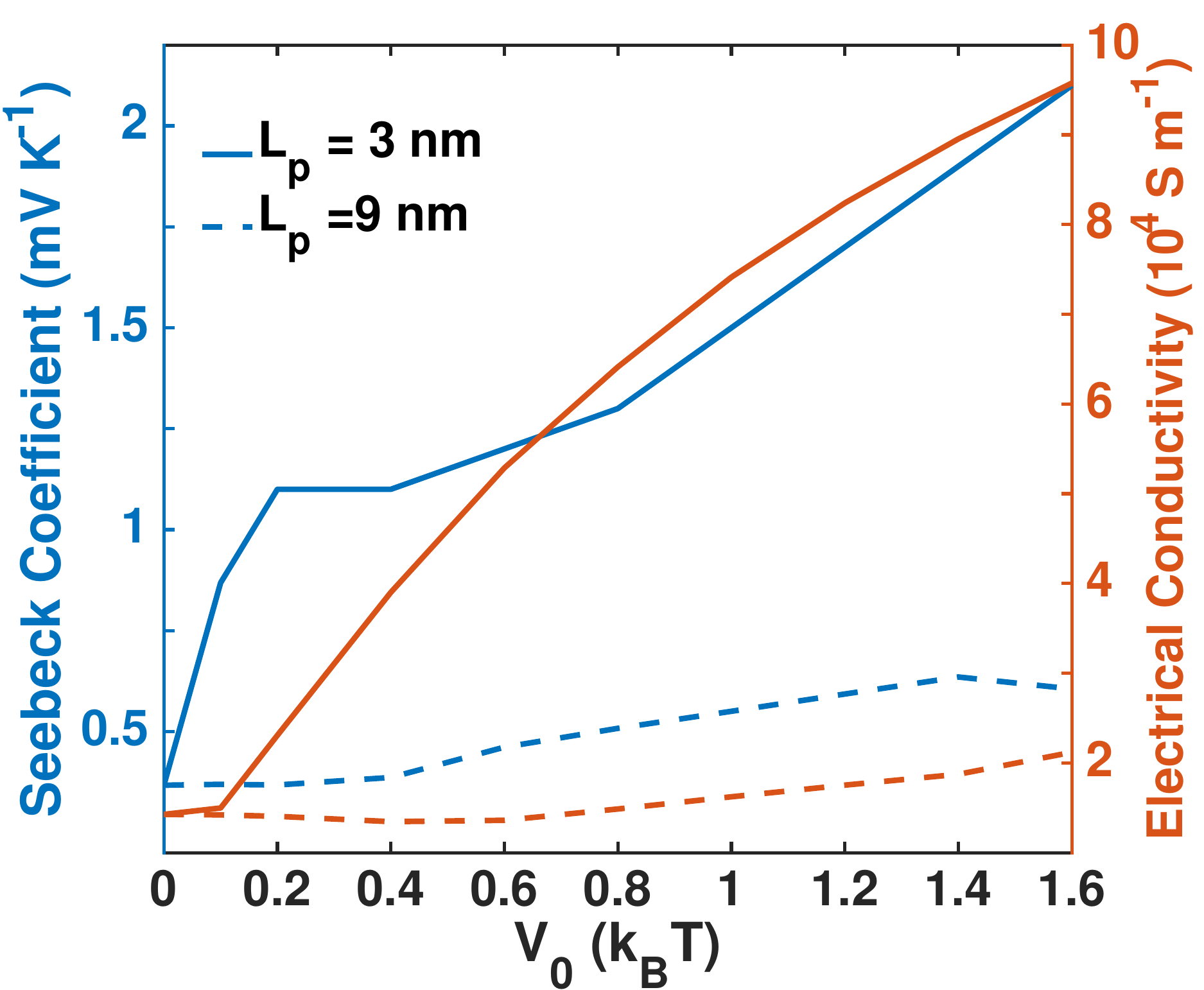}
		\caption{ N$_D$ = 10$^{19}$ cm$^{-3}$}
		\label{3fig6_2}
	\end{subfigure} 
	\begin{subfigure}[b]{0.5\textwidth}
		\centering
		\includegraphics[width=\columnwidth]{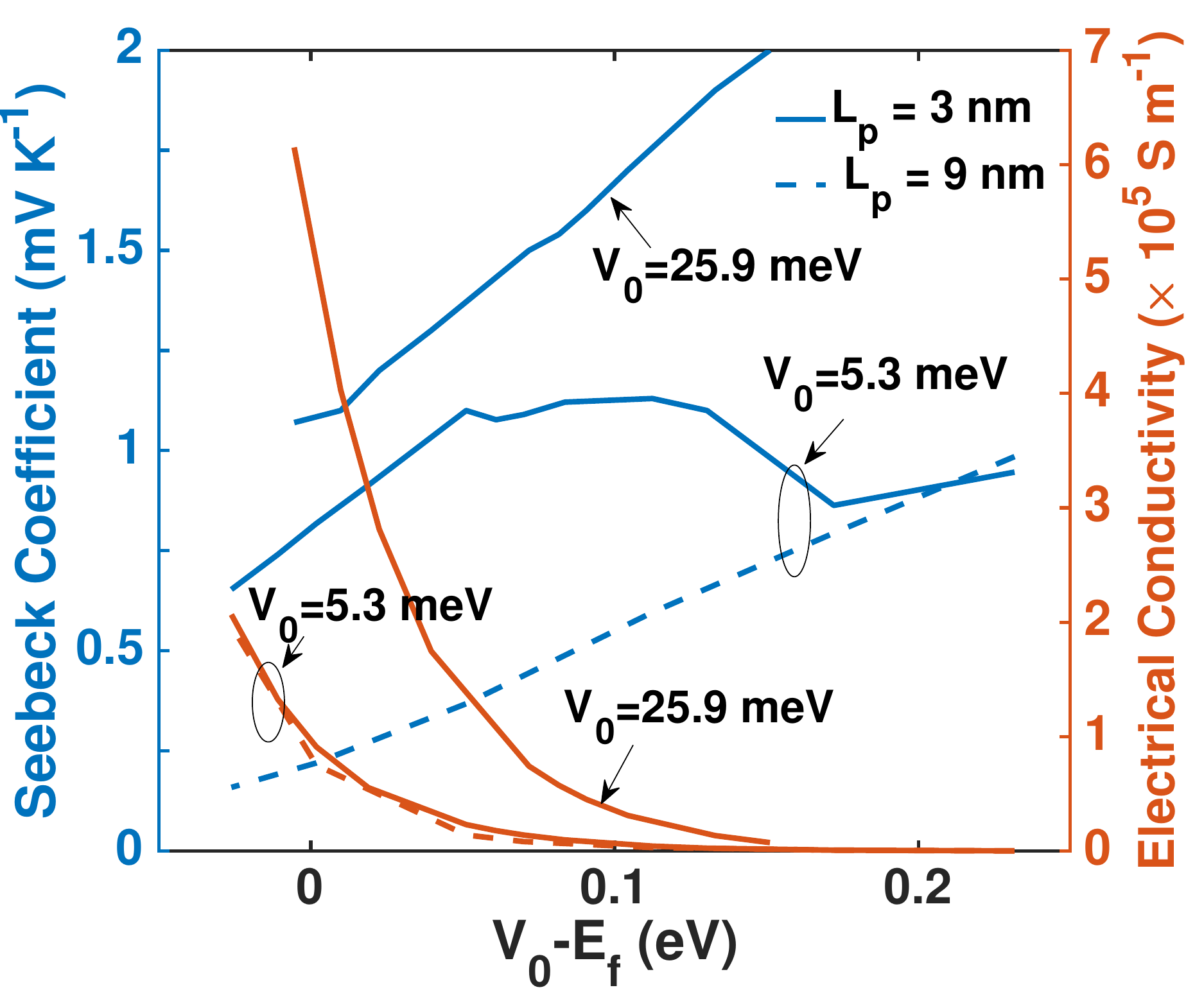}
		\caption{T = 300 K}
		\label{3fig10}
	\end{subfigure}
	\begin{subfigure}[b]{0.5\textwidth}
		\centering
		\captionsetup{justification=centering}
		\includegraphics[width=\columnwidth]{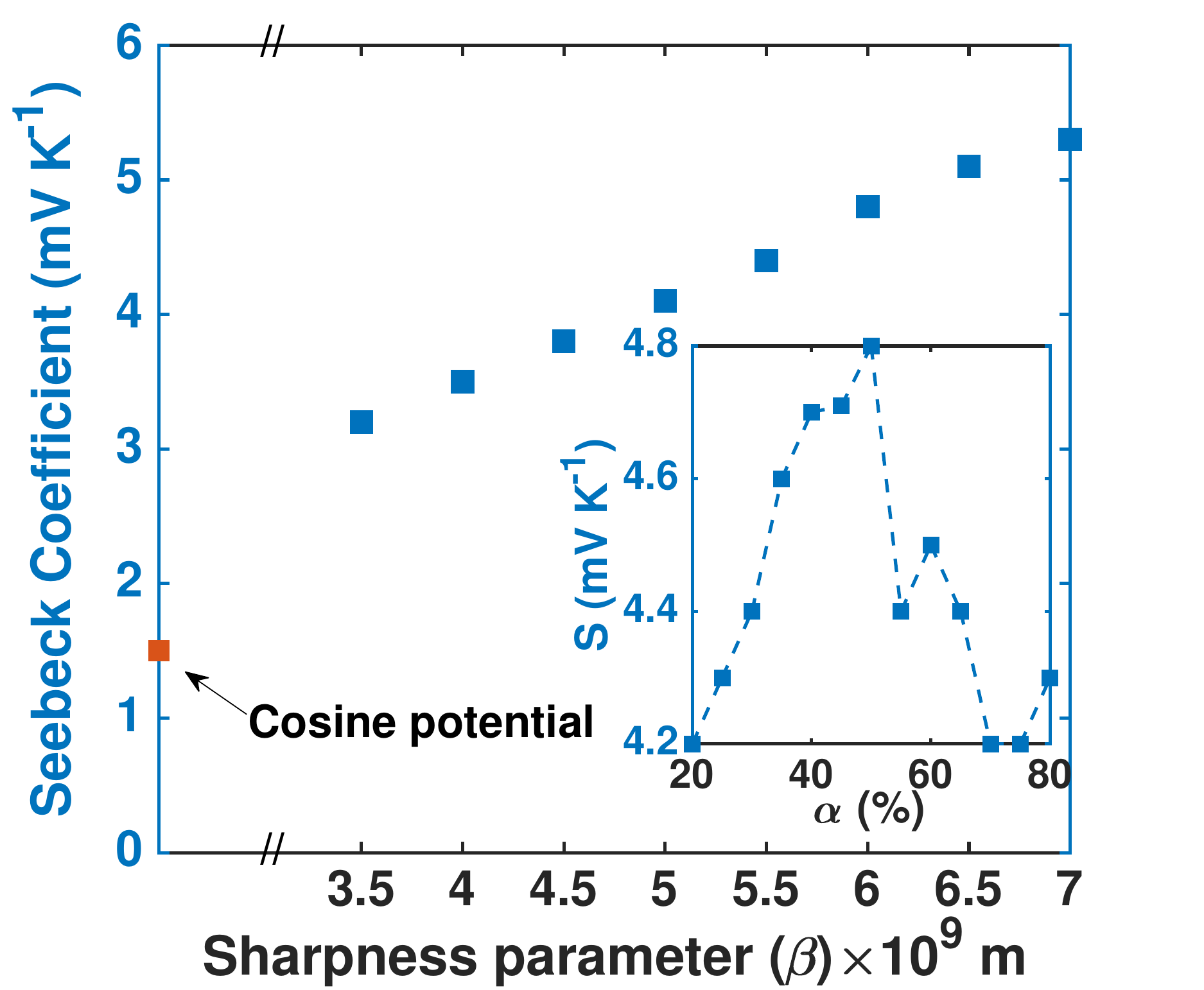}
		\caption{V$_0$ = 1 $k_B$ T and L$_p$ = 3 nm}
		\label{3fig8}
	\end{subfigure}
	\caption{Varying (a) potential period length ($L_p$) (b) peak barrier height ($V_0$) of the potential barriers to understand their effect on thermoelectric performance. (c) Fermi level is tuned to change potential barrier height using the doping of the system. (d) Variation in TE parameters with smoothness of the potential barrier ($\beta$). Inset: Seebeck coefficient is maximum when the barrier width is one-half of the period ($\alpha=50 \%$). A square barrier with $\beta$ = 6$\times$10$^9$ m$^{-1}$ and N$_D$= 10$^{19}$ cm$^{-3}$ at T = 300 K is used. Adapted from \cite{KomminiJPCM2018}}
	\label{3fig6}
\end{figure*}

The solution for the perturbation $g(k)$ to distribution function (Eq.~\ref{Wgi}) in WBTE can now be used to study the thermoelectric behavior of the system. First, to study the effect of potential barrier structure (apart from smoothening), the cosine shaped barriers are used. Introducing potential barriers in highly doped (N$_{D}$ = 10$^{19}$ cm$^{-3}$) silicon with Fermi level of 46 meV below conduction band edge $E_c$ ($E_c$ is set to zero), $g(k)$ is calculated. Cosine-shaped potential barriers with V$_0$ = 1.6 $k_B$T $\cong$ 41 meV at T = 300 K ($k_B$ is the Boltzmann constant), show an decrease in Seebeck coefficient with increase in the period length (as seen in Figure~\ref{3fig6_1}). Increase in period L$_p$ from 3 nm to 9 nm, restricts the tunneling of electrons through the barrier causing a decrease in the Seebeck coefficient. This restriction in tunneling also reduces the electrical conductivity with increase in period length. At L$_p$ = 9 nm, tunneling of carriers vanishes and the enhancement in Seebeck coefficient is a consequence of thermionic emission alone ($\cong$ V$_0$/T $\cong$0.13 meV). Also, by nanostructuring (confining the structure along the directions apart from the transport direction) in the system, we observe a modest decrease in both the Seebeck coefficient and electrical conductivity due to the carrier scattering at the boundaries. Nonetheless, the reduction in thermal conductivity due to nanostructuring produces a significant boost to the overall TE performance.

When the period length is kept constant, increasing the potential barrier height (V$_0$) results in a simultaneous increase in electrical conductivity and Seebeck coefficient, as shown in Figure~\ref{3fig6_2}. Even though the increase in Seebeck coefficient can be attributed to energy filtering, the simultaneous increase in electrical conductivity is a consequence of higher mobility electrons that drive most of the transport. The increase in mobility of electrons causes the conductivity increase with the potential barrier height. Again at wider period lengths (L$_p$ = 9 nm), lack of tunneling and the presence of complete diffusive transport, transport occurs only through thermionic emission. It should be noted that irrespective of the potential barrier height, the uniform doping concentration in the system maintains the number of carriers available. 

Apart from increasing the barrier height, another way of tuning the barrier seen by the carriers is by varying the position of Fermi level. The position of Fermi level E$_F$ is controlled by doping and it strongly influences the Seebeck coefficient, since the Seebeck is defined as the average energy of electrons that participate in the transport relative to the Fermi level. As shown in Figure~\ref{3fig10}, the system acts as bulk/no barrier structure when L$_p$ = 9 nm, replicating the usual TE behavior (an increase in Seebeck coefficient and decrease in conductivity, with higher difference in V$_0$ and E$_f$). Moving Fermi level towards the conduction band by doping the material gives an initial increase in Seebeck coefficient due to combined effect of filtering and an increase in carrier availability. A peak in Seebeck coefficient can be observed when V$_0$-E$_f$ $\cong$ 0.14 eV, which is consistent with previous studies\cite{Neophytou2013}. It is followed by a reduction as the effective barrier height encountered by the carriers decreases and increase in the number of low-energy carriers that can overcome the barrier. An increase in doping results in higher electrical conductivity, which increases the overall power factor at higher doping conditions(smaller V$_0$-E$_f$). 

To study the effect of barrier smoothness, the parameter $\beta$ that controls the smoothness of a square potential is varied (Eq.~\ref{gen_v}). As shown in Figure~\ref{3fig8}, with increase in $\beta$, Seebeck coefficient increases due to the additional quantum reflections that are introduced in the quantum operator (as in Eq.~\ref{fwgen}). These additional reflections at the barrier helps to better filter the carriers compared to smooth barriers. The strength of these reflections increases with the weight for quantum evolution term, which itself is a function of $\beta$. Further, shape parameters like the duty cycle (ratio of potential barrier and the period) also effect the TE performance. Even for sharper barriers (higher $\beta$), Seebeck coefficient is maximum when the barriers are symmetric (50\% duty cycle) as shown in the inset of Figure~\ref{3fig8}.

\subsubsection {Non-equilibrium Green's functions}
The ability to simulate electron-electron correlations in the time domain is a central feature of the non-equilibrium Greens's function (NEGF) formalism, which allows it to capture and simulate many-body quantum effects in nanoscale devices. A comprehensive review on nanoscale device modeling using NEGF was published by Datta ~\cite{DattaSM00}. Here we provide a brief overview of the technique before delving into its recent applications in studying transport in RTDs and thermoelectric devices. The equations of motion in the NEGF formalism in the steady state are given by the Dyson and Keldysh relations~\cite{JauhoBook08,JamesJPC71,Datta1995} as
\begin{equation*}
\hat{G}^R(E)=[E\hat{I}-\hat{H_0}-\hat{V_c}-\hat{\Sigma}^R(E)]^{-1}
\end{equation*}
\begin{equation*}
\hat{G}^<(E)= \hat{G}^R(E)\hat{\Sigma}^<(E)\hat{G}^A(E)
\end{equation*}
\noindent where $\hat{V_c}$ represents the coherently treated interactions that include the mean-field Coulomb potential $\hat{V_c}^{m-f}$ and the electronic disordered part $\hat{V_e}^{rand}$. The spectral self-energy used to treat the incoherent interactions is expressed as
\begin{equation}
\hat{\Gamma}=i(\hat{\Sigma}^R-\hat{\Sigma}^A)=i(\hat{\Sigma}^<-\hat{\Sigma}^>),
\end{equation}
where the retarded self-energy used to include the electron-phonon interactions is defined as
\begin{equation}
\hat{\Sigma}^R(t)=\Theta(t)(\hat{\Sigma}^<-\hat{\Sigma}^>).
\end{equation}
The expectation values of the observables and current operator are calculated using the relation between density matrix and Green's functions. 

\begin{figure}[ht!]
	\centering
	\includegraphics[width=\columnwidth]{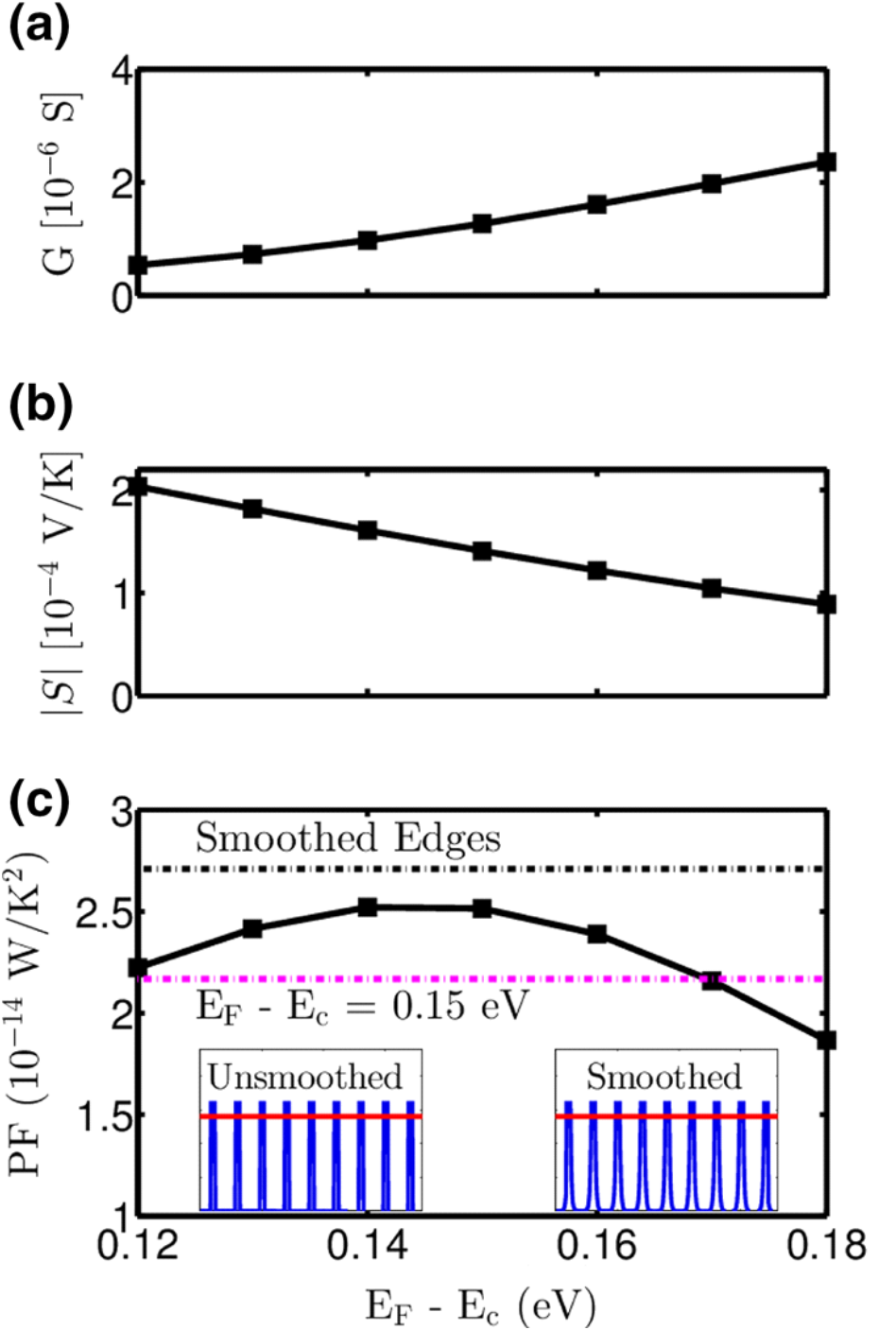}
	\caption{ The thermoelectric performance of a 1D superlattice material modeled using NEGF, similar to the superlattice structure discussed in the previous section. The thermoelectric parameters are plotted as a function of the Fermi level position E$_F$ with respect to the conduction band edge E$_c$ and a barrier height of 0.16 eV is used. a) The electrical conductance, b) the Seebeck coefficient, and c) the power factor. The optimal power factor of the bulk thermoelectric is indicated by the magenta-dashed line, the optimal power factor of a superlattice with smoothened barriers by the black dashed line and compared with the power factor of a unsmoothed barriers in (c). The potential profiles in the superlattice simulated are shown in insets of (c). Reprinted with permission from \cite{Neophytou2013}. Copyright (2016).}
	\label{fig:fig8}
\end{figure}

The NEGF technique has found applications in understanding the impact of quantum effects on electronic and thermoelectric transport, as well as finding ways to further improve TE properties through confinement in heterostructures. The effect of confinement on the thermoelectric performance in Si/Ge/Si superlattices was simulated using NEGF. Strain-induced energy splitting in Si/Ge/Si superlattices is shown to improve power factor by four orders in magnitude~\cite{Bulusu07}. However, the gains in TE performance expected in such structures was shown to be limited by the reduction in conductivity of superlattices with thin barriers~\cite{Bulusu08}. Quantum wire superlattices with lateral confinement were also studied with NEGF, while including the scattering processes due to electron-phonon couplings, phonon anharmonicity, charged impurities, surface and interface roughness and alloy disorder~\cite{Grange2014}. Thermionic emission and tunneling of carriers in periodic superlattices were simulated in the NEGF formalism~\cite{Neophytou2013,Neophytou2016}, and their effect on TE properties has been found to improve the power factor, as shown in Figure \ref{fig:fig8}. These studies also investigated the optimum size and shape of the potential barriers in Si-based superlattices in order to further improve TE performance. They found that hierarchically designed materials, containing heterostructures at varying scales from the atomic to the microscopic, can significantly impact transport and improve the TE power factor~\cite{VargiamidisPRB19}. Double-barrier RTDs were simulated in the NEGF formalism, with self-consistent treatment of inelastic scattering, to show the enhanced valley current which is a consequence of enhanced occupation of resonant state ~\cite{LakeDuttaPRB92}.
   
\section{Conclusions}

We have reviewed recent progress in understanding electronic and thermoelectric transport in heterostructures of 2D and 3D materials. Transport in both lateral and vertical heterostructures is strongly influenced by the band alignment between the two materials while the transmission of carriers between them depends on the simultaneous momentum and energy conservation in the transmission process. Consequently, homojunctions and, to a lesser degree, heterojunctions in 2D materials have conductances that strongly depend on the angle of rotational mismatch between the two adjacent materials in the heterostructure. Thermoelectric properties, including the Seebeck coefficients, are enhanced through heterostructuring in both the 2D and 3D case and offer promise for future on-chip waste-heat energy harvesting applications. Understanding quantum transport in 3D heterostructures and superlattices requires sophisticated modeling based on the Wigner or NEGF approaches. Open problems include simultaneously capturing quantum effects such as tunneling and semi-classical processes such as phonon and impurity scattering, as well as computational optimization of complex multi-layered heterostructures in order to enhance their properties. The latter involves simultaneous materials selection, doping, layer thickness, and, in the 2D case, angle of rotation; this multifaceted computational problem might benefit in the future from machine learning and artificial intelligence techniques in order to identify heterostructures with optimal electronic or thermoelectric properties.

\bibliographystyle{andp2012}
\bibliography{GBmismatchBib,thermoelectric} 

\end{document}